\documentclass[fleqn,usenatbib,useAMS]{mnras}

\usepackage{graphicx}	
\usepackage{amsmath}	
\usepackage{amssymb}	
\usepackage{multicol}        
\usepackage{bm}		
\usepackage{pdflscape}	

\usepackage[T1]{fontenc}
\usepackage{ae,aecompl}

\usepackage{newtxtext,newtxmath}

\newcommand{\noopsort}[1]{}
\newcommand{\ratio}[2]{\text{#1}/\text{#2}}

\title[Diverse material orbiting white dwarfs]{Interpretation and diversity of exoplanetary material orbiting white dwarfs}

\author
[A. Swan et al.]
{Andrew Swan$^{1}$\thanks{E-mail: \href{mailto:a.swan.17@ucl.ac.uk}{a.swan.17@ucl.ac.uk}},
Jay Farihi$^{1}$,
Detlev Koester$^{2}$,
Mark Hollands$^{3}$,
Steven Parsons$^{4}$,
\newauthor
P. Wilson Cauley$^{5}$,
Seth Redfield$^{6}$, and
Boris T. G{\"a}nsicke$^{3}$
\\
$^{1}$Department of Physics~\& Astronomy, University College London, Gower Street, London~WC1E~6BT, UK\\
$^{2}$Institut f\"ur Theoretische Physik und Astrophysik, University of Kiel, D-24098~Kiel, Germany\\
$^{3}$Department of Physics, University of Warwick, Coventry~CV4~7AL, UK\\
$^{4}$Department of Physics and Astronomy, University of Sheffield, Sheffield~S3~7RH, UK\\
$^{5}$Laboratory of Atmospheric and Space Physics, University of Colorado Boulder, Boulder, CO~80309, USA\\
$^{6}$Department of Astronomy and Van Vleck Observatory, Wesleyan University, Middletown, CT~06459, USA\\
}

\date{Accepted XXX. Received YYY; in original form ZZZ}

\pubyear{2019}

\begin{document}
\label{firstpage}
\pagerange{\pageref{firstpage}--\pageref{lastpage}}
\maketitle

\begin{abstract}
Nine metal-polluted white dwarfs are observed with medium-resolution optical spectroscopy, where photospheric abundances are determined and interpreted through comparison against solar system objects. An improved method of making such comparisons is presented that overcomes potential weaknesses of prior analyses, with the numerous sources of error considered to highlight the limitations on interpretation. The stars are inferred to be accreting rocky, volatile-poor asteroidal materials with origins in differentiated bodies, in line with the consensus model. The most heavily polluted star in the sample has 14~metals detected, and appears to be accreting material from a rocky planetesimal, whose composition is mantle-like with a small Fe--Ni core component. Some unusual abundances are present: one star is strongly depleted in Ca, while two others show Na abundances elevated above bulk Earth, speculated either to reflect diversity in the formation conditions of the source material, or to be traces of past accretion events.  Another star shows clear signs that accretion ceased around 5\,Myr ago, causing Mg to dominate the photospheric abundances, as it has the longest diffusion time of the observed elements. Observing such post-accretion systems allows constraints to be placed on models of the accretion process.
\end{abstract}




\section{Introduction}
\label{sectionIntroduction}

White dwarfs are the cooling remnants of intermediate-mass stars, whose main-sequence masses are typically 1--3\,{M\textsubscript{\sun}} \citep{Tremblay2016,Cummings2018}. They are compact objects, Earth-sized but with masses around 0.6\,{M\textsubscript{\sun}}, and thus high surface gravities. They are therefore expected to show spectral lines only of H or He, as metals sink below their atmospheres on time-scales that are orders of magnitudes shorter than their cooling age \citep{Paquette1986, Koester2009}. However, in approximately one-third to one-half of cases that expectation proves incorrect \citep{Koester2014}, with as many as 15~metals observed in the most extreme case \citep{Zuckerman2007}. Heavy elements can be (partly) supported through radiative levitation, though significant work remains before that process is well understood, but in $T_{\text{eff}}\lesssim25\,000$\,K stars its effect is negligible \citep{Barstow2014, Koester2014}. Thus, any photospheric metals must have an external origin, so the stars are said to be polluted. In short-period binary systems, this pollution can come from the wind of the companion star \citep{Debes2006}, but single stars require a distinct explanation. Historically, accretion from the interstellar medium (ISM) was thought to be responsible \citep{Dupuis1993}, but that has been found to be incompatible with observations. He-atmosphere stars have H~abundances orders of magnitude lower than can be accounted for by that mechanism, the predicted accretion rates fail to account for the metal quantities observed, and no correlation has been found between metal pollution and ISM density \citep{Koester2000, Dufour2007,Farihi2010rockyPlanetesimals}.

A compelling body of evidence now points to disrupted asteroidal material as the source of the pollution. In the favoured model, a major planet perturbs the orbits of asteroids so that they pass within the stellar Roche limit, where they are tidally disrupted \citep{Jura2003}. Collisions in the resulting debris disc evolution produce micron-sized dust particles that spiral inwards under Poynting-Robertson drag \citep{Rafikov2011PRdrag} until they sublimate. Viscous evolution of the resulting gas then leads to accretion onto the star
\linebreak 
\citep{Bochkarev2011,Kenyon2017collisions}. Sinking time-scales for metals range from a few days in warm, H-dominated atmospheres to a few million years in cool, He-dominated atmospheres \citep{Dupuis1992, Koester2009}\footnote{See also the note and updated tables at \href{http://www1.astrophysik.uni-kiel.de/~koester/astrophysics/astrophysics.html}{www1.\hspace{0pt}astrophysik.\hspace{0pt}uni-kiel.de/\hspace{0pt}~koester/\hspace{0pt}astrophysics/\hspace{0pt}astrophysics.html}.}. In addition to metal pollution, several strands of evidence support this model, the most important of which are: infrared excesses consistent with circumstellar dust, emission and absorption lines from circumstellar gas, and transits by opaque clouds of material (e.g. \citealt{Jura2009}; \citealt{Melis2012}; \citealt{Vanderburg2015}; \citealt{Cauley2018}). A supply of small bodies and a planet to perturb their orbits are implied \citep{Bonsor2011, Debes2012}, but neither these, nor the reservoirs of cold dust that might accompany an asteroid belt, are yet observed \citep{Farihi2014Alma}. However, the \textit{Gaia} spacecraft is anticipated to be able to detect giant planets around white dwarfs astrometrically by the end of its mission, should they exist in a favourable region of parameter space \citep{Perryman2014}.

Abundances of accreted material can be inferred from photospheric abundances, given a model for the accretion process \citep{Koester2009}. A model for white dwarf atmospheres is also required, and the diffusion processes operating therein are an active area of research \citep{Bauer2018, Cunningham2019}. A few dozen polluted stars have been examined in detail spectroscopically, predominantly showing abundances consistent with rocky, volatile-poor material \citep{Jura2014}. [Note that classification of elements in this study as volatile (C, N, O, and to some extent Na) is by reference to their condensation temperature \citep{Lodders2003}.] In a few instances the presence of water may be inferred, owing to an excess of O beyond that expected in rock-forming minerals \citep{Farihi2013, Raddi2015}, while only one system appears to be accreting volatile-rich material consistent with cometary abundances \citep{Xu2017}. At least some of the material is thought to originate from differentiated bodies \citep{Zuckerman2011, Jura2013, Hollands2018analysis}. Observational signatures of differentiated parent bodies include overabundance (relative to chondrites) of lithophile elements, such as Al, Ca, and Ti, that concentrate in the crust, or material rich in Fe and Ni in a ratio similar to that expected in planetary cores. Spectroscopy of white dwarfs is presently the only empirical method that allows such direct access to the bulk composition of exoplanets.

This paper presents medium-resolution optical spectroscopy of nine polluted white dwarfs. The observations are described in Section~\ref{sectionObservations}. Atmospheric modelling, determination of stellar parameters, derived elemental abundances, and the analysis of those abundances are detailed in Section~\ref{sectionMethods}. Results are presented in Section~\ref{sectionResults}, and discussed in Section~\ref{sectionDiscussion}. Section~\ref{sectionSummary} provides a short summary of the results and conclusions.

\section{Observations}
\label{sectionObservations}

The sample was drawn from known polluted white dwarfs in the southern sky \citep{Friedrich2000, Koester2005, Subasavage2007}. Stars are referred to throughout this study by their WD~designation as catalogued (abbreviated B1950.0 coordinates; \citealt{McCook1999}).

Measurements were taken using the {X-shooter} echelle spectrograph \citep{Vernet2011} on the ESO UT2~Kueyen telescope on Cerro Paranal. Observations were made in service mode on dates between 2011~April~3 and August~30. The instrument provides simultaneous optical and near-infrared coverage across three independent arms. Wavelength coverages are roughly {3000--5500\,\AA}, {5500--10\,000\,\AA}, and {10\,000--25\,000\,\AA} in the UVB, VIS, and NIR arms, respectively (this ordering is maintained throughout the paragraph). Slit widths were set to $0\farcs5$, $0\farcs4$, and $0\farcs6$, for nominal resolving powers $R=\uplambda/\upDelta\uplambda$ of 9100, 17\,400, and 8100. Exposure times were typically 1475\,s, 1420\,s, and 600\,s, with total integration times of between 1\,h and 5\,h depending on the target brightness. The UVB and VIS detectors were employed in high-gain, slow-readout mode, with no binning in the spatial direction, and two-pixel binning in the dispersion direction. Observations were taken in nodding mode with a {5\arcsec}~throw.

The data were processed using version~2.9.3 of the {X-shooter} pipeline, following standard procedures. Each spectrum was extracted and wavelength-calibrated using default settings, with the exception that the position of the target spectrum was localised on the slit using Gaussian fitting and allowing for some tilt. The spectra were sensitivity-calibrated using spectrophotometric standard star observations taken on the same night.

\begin{figure*}
 \centering
 \includegraphics[width=\textwidth]{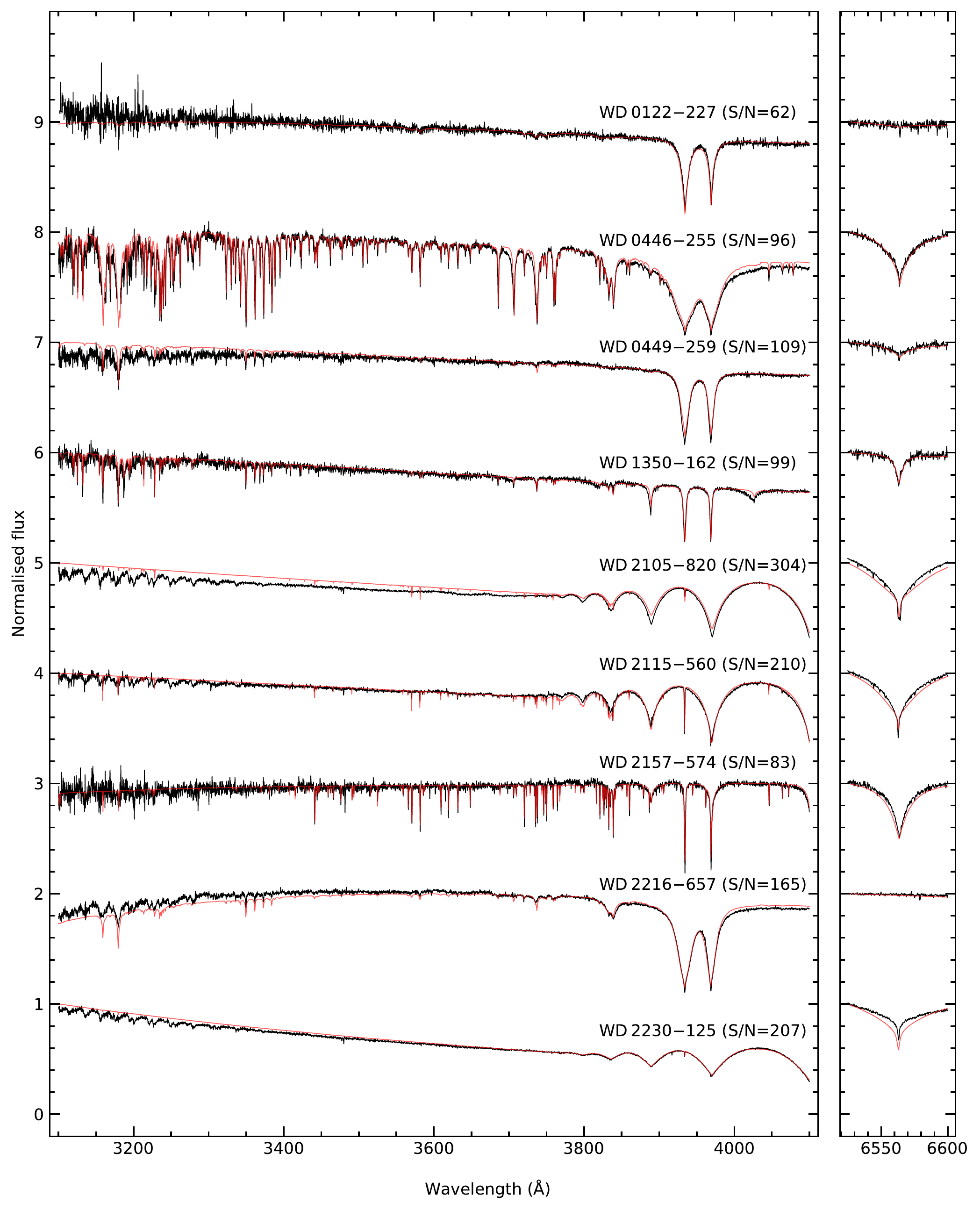}
 \caption{Extracts from X-shooter spectra (black) for all stars in the sample, overlaid with models (red). Fluxes are normalised within each section, and offset vertically. S/N is quoted at 4000\AA.}
 \label{figureSpectra}
\end{figure*}

Signal-to-noise (S/N) ratios achieved were in the ranges 60--285 in the UVB arm and 30--200 in the VIS arm. The data from the NIR arm had $\text{S/N}<10$ and were not used in this analysis. Inspection of pixel-domain power spectra reveals periodicities, particularly in the UVB data. As they are present across most of the sample, they are likely to be fringing in the instrument light path. While effective S/N may be reduced, and some weaker lines may be obscured, the effect is not otherwise considered to significantly impact abundance determination. Spectra for all stars are shown in Figure~\ref{figureSpectra}, overlaid with the models described in Section~\ref{sectionMethods}, and S/N at 4000\AA.

For 0446$-$255, 0449$-$259, 1350$-$162, and 2230$-$125, the {X-shooter} spectra were supplemented with ancillary data from the HIRES echelle spectrograph on the Keck~I telescope on Mauna Kea \citep{Vogt1994}. Observations were made on 2017~June~27 and 2018~September~6. The C5 decker was used, providing a slit width of 1\farcs148, and a resolving power $R=\uplambda/\upDelta\uplambda=34\,000$. Wavelength coverage is 3130--5990\,\AA, and $\text{S/N}\approx10\text{--}20$ was achieved. Standard data reduction procedures were followed, using the \textsc{hires redux} package. Spectra extracted from individual exposures were co-added, and a median filter routine was applied to correct pixels affected by cosmic rays.

An ultraviolet spectrum of 2230$-125$ was taken with the COS instrument on board the \textit{Hubble Space Telescope} (\textit{HST}) on 2017~May~9, using the G130M grating at a central wavelength of 1291\,\AA, providing resolving power of 12\,000--16\,000. The exposure time was 2180\,s, achieving $\text{S/N}\approx24$. The observations were obtained in a single \textit{HST} orbit, using all four FP-POS settings to minimize the effect of fixed pattern noise. The data were retrieved from the \textit{HST} archive and reduced with \textsc{calcos}~3.3.4.

\section{Methods}
\label{sectionMethods}

The spectra are analysed by fitting model white dwarf atmospheres to determine stellar parameters and element abundances, as described in Section~\ref{subsectionAtmospheresAndAbundances}. The treatment of uncertainties is set out in Section~\ref{subsectionUncertainties}. The abundances are then compared to solar system materials using a $\upchi^{2}$~goodness-of-fit test, as described in Section~\ref{subsectionChiSquaredComparison}.

\subsection{Model atmospheres and abundance determination}
\label{subsectionAtmospheresAndAbundances}

The physics described in \citet{Koester2010} are used to construct model atmospheres and synthetic spectra, taking atomic line data from the NIST and VALD3 databases \citep{Piskunov1995,Ryabchikova2015,Kramida2018}. For the H-atmosphere stars, effective temperature and surface gravity are determined by fitting the spectra to a grid of pure H atmospheres, i.e. using only the Balmer lines. The $\log{g}$ values are then refined using parallaxes and photometry from \textit{Gaia} Data Release~2 \citep{Prusti2016,Brown2018}. To confirm that reliable fits can be obtained without including metals, three model atmospheres are constructed for a given star as follows: (i)~no metals, (ii)~all observed elements, and (iii)~all elements including upper limits. The models with and without metals are found to exhibit no morphological differences at the appropriate spectral scale. Once determined, $T_{\text{eff}}$ and $\log{g}$ are held constant while other species are introduced to the model atmospheres, and the metal abundances are varied to find the best fit between the lines identified in the spectra and the models.

A two-stage iterative approach is taken for the He-atmosphere stars, whose spectra are less sensitive to surface gravity. First, stellar parameters are determined while abundances are held constant, and then abundances are determined while stellar parameters are held constant. The whole procedure is iterated until the fit converges. Radius and $T_{\text{eff}}$ are found by fitting to photometry with the \textit{Gaia} parallax included as a prior parameter. Where available, photometry is taken from Pan-STARRS, SkyMapper, and \textit{GALEX} \citep{Martin2005, Chambers2016, Wolf2018}. Surface gravity is derived by interpolating over a grid of evolutionary models\footnote{\href{http://www.astro.umontreal.ca/~bergeron/CoolingModels}{www.astro.umontreal.ca/$\sim$bergeron/CoolingModels/} (thin H sequence).} \citep{Fontaine2001}. Abundances are determined by fitting to synthetic spectra.

\begin{figure}
 \includegraphics[width=\columnwidth]{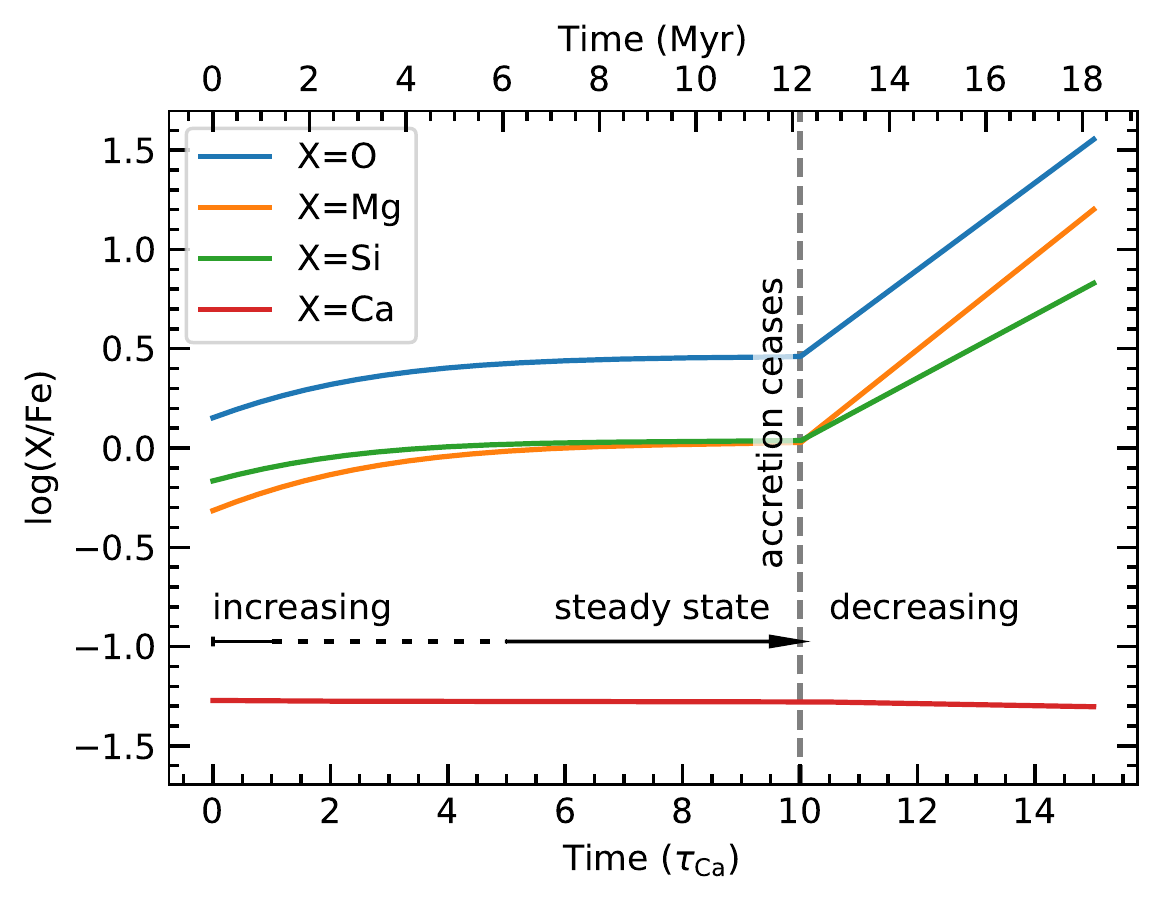}
 \caption{Illustration of the three phases in the evolution of photospheric abundance ratios, where accretion of bulk-Earth material proceeds at a constant rate until it ceases abruptly.}
 \label{figureThreePhases}
\end{figure}

Abundances of the accreted debris must be inferred from the photospheric abundances, as the former are modified by diffusion processes in the stellar atmosphere. By assuming a constant rate of accretion, three regimes can be identified in the evolution of photospheric abundances \citep{Koester2009}. There is no consensus on the terminology for these phases, but \textit{increasing}, \textit{steady-state}, and \textit{decreasing} are used in this study as they reflect the behaviour of the abundances, but do not presume anything about the history (unlike e.g. \textit{early} and \textit{late}). Fig.~\ref{figureThreePhases} provides an illustration of how observed abundance ratios change during and after an accretion event. Material with bulk-Earth composition is assumed to accrete onto a He-dominated star ($T=13$\,000\,K; $\log{g}=8.0$) at a constant rate, ceasing after 10~Ca-sinking time-scales\footnote{The sinking time for Ca, $\uptau_{\text{Ca}}$, is used throughout as a benchmark.} have elapsed. Initially the photospheric ratios are those of the accreted material (the increasing phase), but after several sinking time-scales they asymptotically approach their steady-state values. Metal-to-metal ratios do not differ dramatically between these two phases, but once accretion ceases (the decreasing phase) these ratios diverge exponentially, departing by an order of magnitude from those of the accreted debris within a few sinking time-scales. Note that in this example, the sinking time-scales of Ca and Fe are similar, thus the change in the $\ratio{Ca}{Fe}$ ratio is minimal.

The short sinking time-scales in H-dominated atmospheres make it likely that accretion will be in a steady state, especially for $T_{\text{eff}}>10\,000$\,K. Sinking time-scales in He-dominated atmospheres are much longer, and in some cases a few million years must elapse before a steady state can be attained. Disc lifetimes are uncertain, though possibly $\sim10^5$\,yr, and therefore it is unclear which accretion phase should be expected at He-dominated stars \citep{Girven2012, Bergfors2014}.

\subsection{Uncertainties}
\label{subsectionUncertainties}

Errors determined for $T_{\text{eff}}$ and $\log{g}$ during the fitting procedure are judged to be formal for the He-atmosphere stars, and are reported here. However, fitting errors for the H-atmosphere stars (typically $\pm 10$\,K and $\pm 0.001$\,dex respectively) are purely statistical and severely underestimate the true uncertainties. Thus, the reported values include additional correlated Gaussian contributions of 3.5~per~cent for $T_{\text{eff}}$ and 0.08\,dex for $\log{g}$ to approximate measurement errors. Systematic errors undoubtedly arise in the modelling, and therefore additional uncorrelated errors of 2~per~cent for $T_{\text{eff}}$ and 0.08 (H~atm.) or 0.25 (He~atm.) for $\log{g}$ are introduced during subsequent analysis. These additional error components are averages of those reported in analyses of much larger samples of stars with the models used here \citep{Koester2014, Koester2015}, and are conservative as the data in those studies have lower S/N. Absolute abundances are tightly correlated with $T_{\text{eff}}$, while relative abundances are negligibly affected by changes in stellar parameters on the scale of the uncertainties here. Therefore an additional uncertainty of 0.2~dex is imposed on the abundances, correlated with $T_{\text{eff}}$. That uncertainty propagates through to derived quantities such as the accretion rate, but it does not affect abundance ratios. It must be emphasized that the approach presented here is an approximation based on best estimates of the uncertainties; the unknown systematic errors prevent a more rigorous approach. Error propagation beyond the model fitting is achieved using a Monte Carlo method. Calculated quantities are reported as the median of the distribution, with the 68~per~cent of the distribution centred thereon treated as a $1\upsigma$ uncertainty.

\subsection{Comparison with solar system material}
\label{subsectionChiSquaredComparison}

Taking a similar approach to \citet{Xu2013}, an attempt is made to quantify the similarity between the accreted material and some solar system objects, using a goodness-of-fit test. Comparisons are made with bulk silicate Earth\footnote{Also known as the primitive mantle, representing the composition of the upper layers of the Earth after the core separated. It is abbreviated PRIMA in the Figures.}, bulk Earth, its core, its crust, solar abundances, and comets 67P/Churyumov--Gerasimenko (\mbox{67P/C--G}) and 1P/Halley \citep{Jessberger1988,Lodders1998,Rudnick2003,Bardyn2017,Wang2018}. A database of meteorite abundance analyses is also used \citep{Nittler2004}; weighted means and standard deviations for each meteorite group are calculated. Most measurements are of whole-rock samples (which measure bulk composition), and those that are not, or have few elements reported, are down-weighted. Meteorite groups are considered if they had measurements for all of Na, Mg, Al, Si, Ca, Ti, Cr, Fe, and Ni since those elements are present in the stars studied here. Pallasites are also included as comparisons, although they lack determinations for Na and Ti.

It must be emphasized that while the meteorite database was assembled primarily from the peer-reviewed literature, no judgement was made by its authors on the quality of the data, as the priority was inclusivity. It is also important to note that meteorites display significant variation within classes, and even within individual objects. For example, the database contains 13~meteorites in the mesosiderite class that have 10 or more separate determinations of Mg, Si and Fe fractions. The standard deviations of $\log{(\ratio{Fe}{Si})}$ ratios for those meteorites range from 0.18 to 0.64. Thus, a variety of compositions may be consistent with a given meteorite class.

The comparison set is dominated by the meteorite groups, whereas the true range of compositions of exoplanetary material could in principle be far wider \citep{Bond2010}. However, the meteorites have been scattered towards the inner solar system, as must also be the case for the material polluting white dwarfs, and the compositions inferred for the majority of systems studied to date do indeed appear rocky \citep{Jura2014}.

Abundances in the form of mass fractions are used for comparison. Due to the heterogeneous sets of measured abundances across the sample, the fractions are measured with respect to only the observed metals in each star. Abundance uncertainties for solar system objects are combined with the observation uncertainties. A reduced~$\upchi^{2}$ statistic (referred to simply as $\upchi^{2}$ hereafter) is calculated:

\begin{equation}
\label{equationChiSquared}
\upchi^{2}=\frac{1}{N-1}\sum_{\text{i}=1}^{N}\frac{(\text{logit}(Z_{\text{i,s}})-\text{logit}(Z_{\text{i,c}}))^2}{\upsigma^{2}}
\end{equation}

\noindent where $N$ is the number of metals observed (for $N-1$ degrees of freedom), $Z_{\text{i,s}}$ and $Z_{\text{i,c}}$ are the mass fractions of each element in the star and comparison respectively, and $\upsigma$ is the combined uncertainty. For a proportion $p$, where $0<p<1$, the logit function takes the form:

\begin{equation}
\label{equationLogit}
\text{logit}\,p=\log{\left(\frac{p}{1-p}\right)}
\end{equation}

More detail on the logit transform is given in Appendix~\ref{appendixChiSquared}, with notes on the interpretation of the $\upchi^{2}$ results and the various sources of uncertainty.

\section{Results}
\label{sectionResults}

Table~\ref{tableResults} lists the stellar parameters, mixing layer depths ($M_{\text{cvz}}$), and diffusion timescales for Ca, that are derived from the white dwarf atmosphere model \citep{Koester2010}. Note that the mixing layer is used here to mean the surface convection zone, when present, or the region with optical depth $\uptau_{\text{R}}\leq5$ (relevant only to the radiative atmosphere of 2230$-$125). Also given are the photospheric abundances, with inferred minimum accretion rates ($\dot{M}_{\text{Z}}$) and metal masses ($M_{\text{Z}}$) in the mixing layer. Distances found by inverting \textit{Gaia} parallaxes are given for reference. Only detected elements contribute to the calculated $\dot{M}_{\text{Z}}$ and $M_{\text{Z}}$. These values are therefore lower limits, particularly where the dominant rock-forming elements O and Si are not detected. $\dot{M}_{\text{Z}}$ and $M_{\text{Z}}$ increase at most by a factor of two, i.e. within the uncertainties, if missing elements are extrapolated from bulk Earth abundances relative to Mg. The reported $\dot{M}_{\text{Z}}$ assumes steady-state accretion, and so for He-dominated stars should be interpreted as a time-averaged, not instantaneous rate. Note that unless otherwise specified, element ratios are quoted by mass throughout the paper, assuming steady-state accretion.

The range of $M_{\text{Z}}$ is consistent with an origin in minor planets or fragments thereof. The star with the highest $M_{\text{Z}}$ (0446$-255$) also has the highest atmospheric H mass amongst the He-dominated stars here. Its $M_{\text{Z}}=10^{23}$\,g is comparable to 10~Hygiea or Enceladus, and at 2\,g\,cm\textsuperscript{-3} would form a sphere of radius 200\,km. Values of $\dot{M}_{\text{Z}}$ among the sample are generally typical of the population (\citealt{Farihi2016}, cf.~fig.~10). While $\dot{M}_{\text{Z}}$ for 2230$-125$ is two orders of magnitude lower than usually found for similar-temperature stars, it is within the range for stars sensitively observed by \textit{HST}~COS \citep{Koester2014}.

The high S/N data reveal multiple metals at each star in the sample. The median number of detections is five, reaching up to 14. All the stars have Ca lines, and most display Mg and Fe, but other elements are detected less uniformly. As is typical of polluted white dwarfs, detections are predominantly of refractory elements, though three measurements of moderately-volatile Na are achieved. Unfortunately, the poorly-constrained volatile abundances allow little insight into the ice and organic content of the accreted materials. By contrast, rock-forming elements are well represented, and indeed all the stars appear to be accreting material originating in rocky bodies. Beyond this, no single theme unifies the sample, which displays considerable diversity when examined in detail. Some unusual features are present: compared to bulk Earth, 2105$-$820 is markedly Ca-deficient, while 0449$-$249 and 1350$-$162 are enhanced in Na, and 2216$-$657 is enhanced in Mg.

The accreted material is now examined using the three-phase model of photospheric abundance evolution described in Section~\ref{subsectionAtmospheresAndAbundances}. A qualitative overview is given in Section~\ref{subsectionAbundanceRatios} by reference to selected element ratios. A quantitative analysis follows in Section~\ref{subsectionChiSquaredResults}, where the $\upchi^{2}$~procedure described in Section~\ref{subsectionChiSquaredComparison} is used to compare the material accreted by each star against solar system compositions.

\subsection{Abundance ratios}
\label{subsectionAbundanceRatios}

\begin{figure}
 \includegraphics[width=\columnwidth]{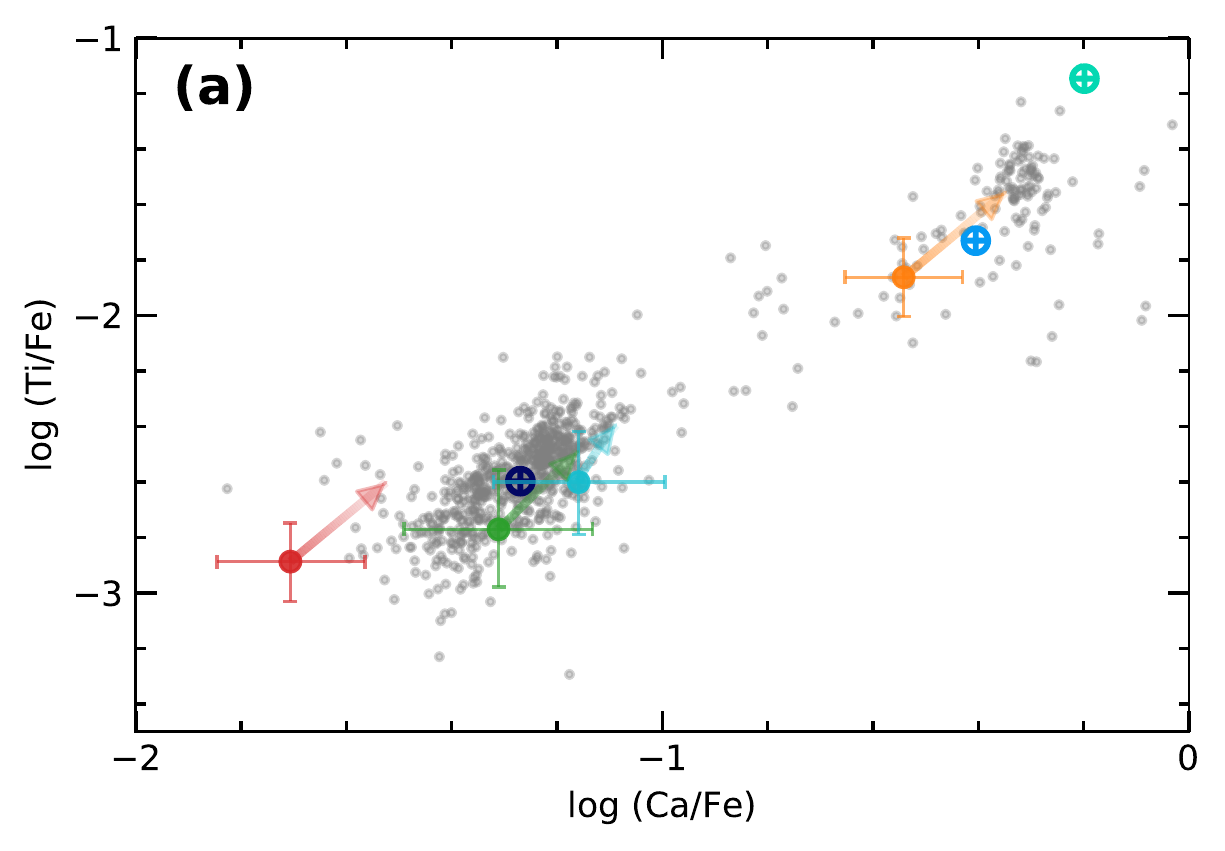}
 \includegraphics[width=\columnwidth]{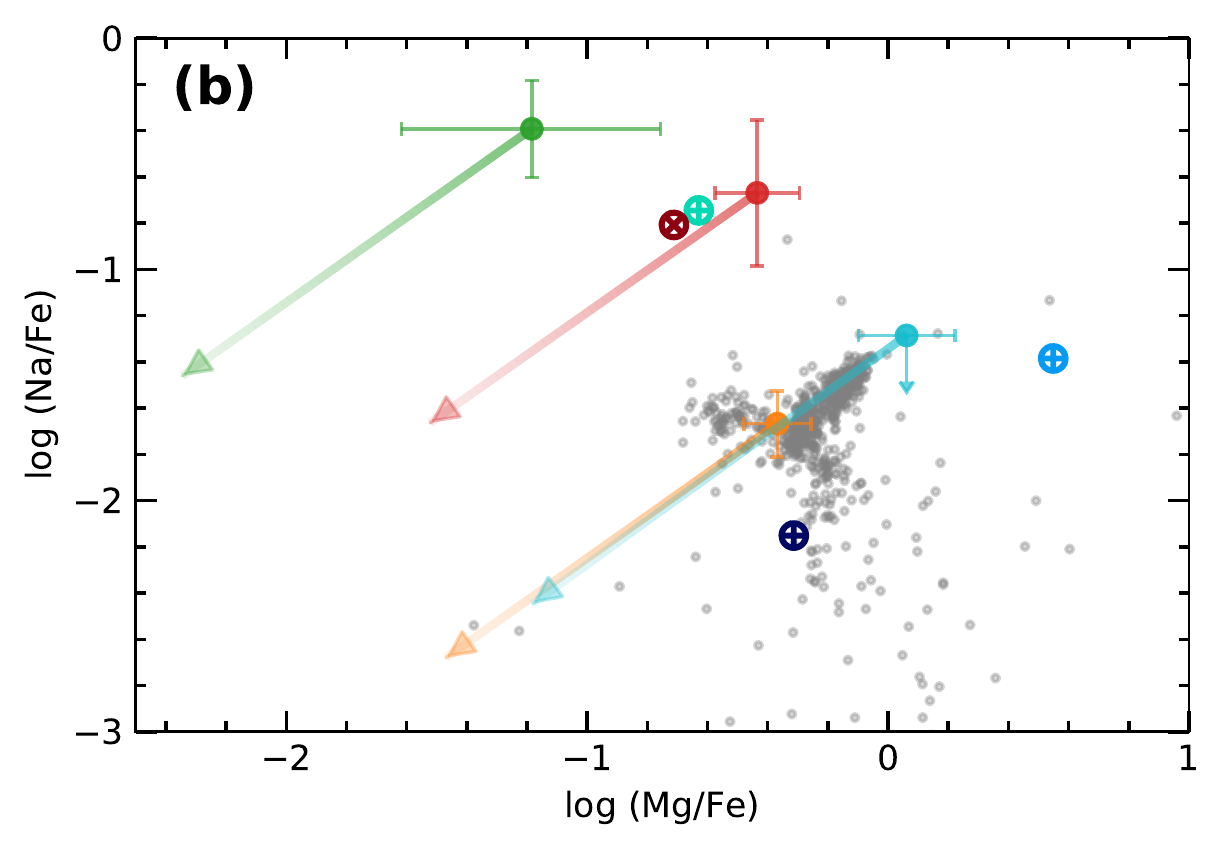}
 \includegraphics[width=\columnwidth]{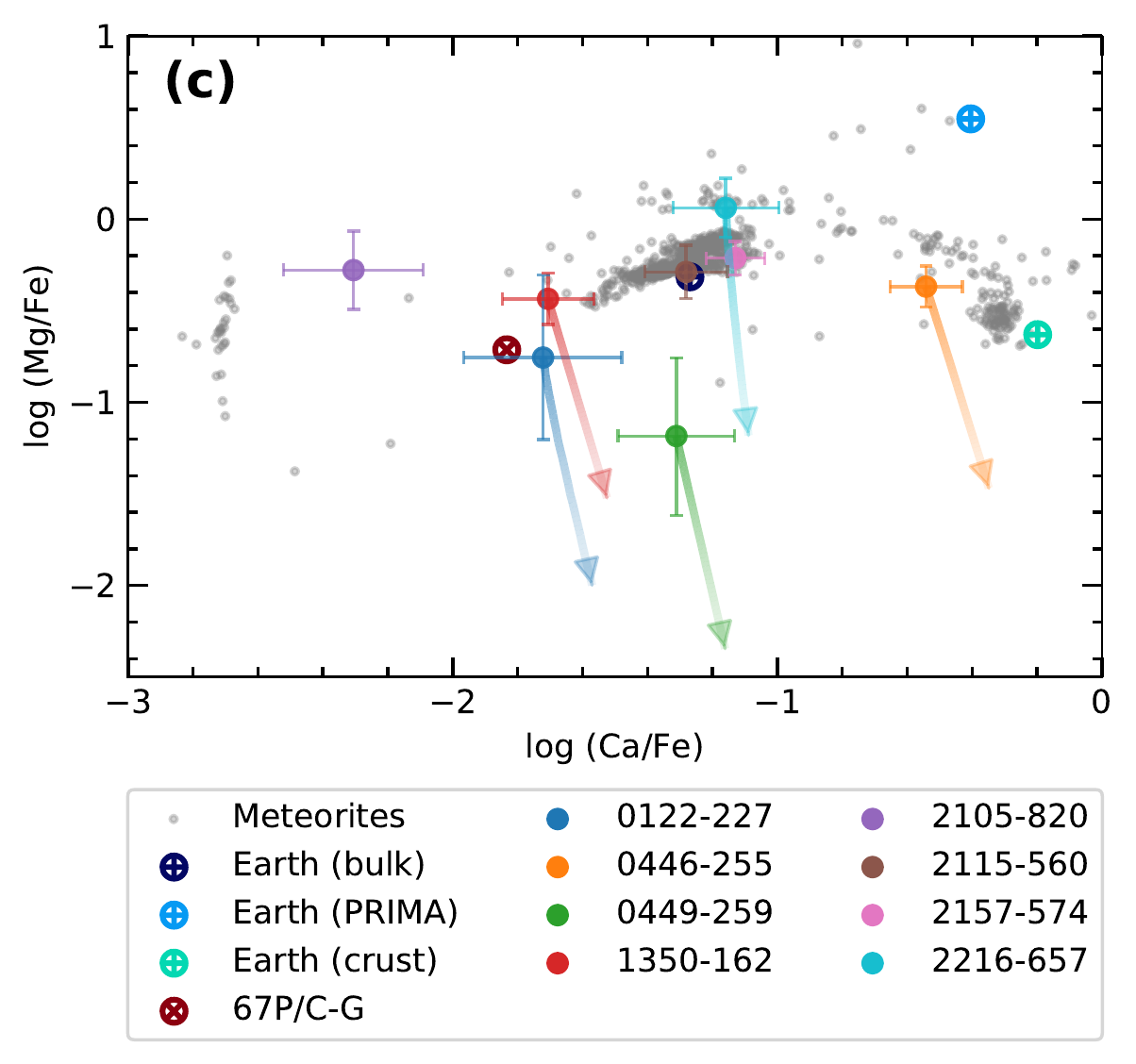}
 \caption{Log mass abundance ratios. Filled circles and error bars show the inferred ratios in the accreted material, assuming a steady state. The arrows show how that inference changes if the system is now in the decreasing phase: accreted abundance ratios are effectively traced back in time toward the arrowhead, where accretion ceased 5\,$\uptau_{\text{Ca}}$ ago. Arrows are shown only for He-atmosphere stars. Panel~(a) plots lithophiles Ca and Ti relative to siderophile Fe. Panel~(b) highlights unusual Na abundances. Panel~(c) plots elements common to most of the sample.}
 \label{figureElementRatios}
\end{figure}

Selected elemental abundance ratios for the accreted material are presented in Fig.~\ref{figureElementRatios}. Coloured points represent the observed stars, assuming steady-state accretion. Results for the increasing phase would appear similar on these logarithmic scales, as the abundance ratios differ by less than a factor of three between the increasing and steady-state phases. The arrows show how the inferred ratios evolve in the decreasing phase, effectively looking back in time to the compositions required (arrow head) to produce the observed abundances (arrow tail). The arrow length indicates five sinking times since accretion ended. Arrows are shown only for He-atmosphere stars, where accretion is more likely to have departed from steady state. Earth abundances are also plotted for comparison, as are \mbox{67P/C--G} and individual meteorites from the database used in this study \citep{Nittler2004}.

In Fig.~\ref{figureElementRatios}(a), all four stars with measurements lie near the same line of constant $\ratio{Ca}{Ti}$ as most solar system objects. Three occupy the same region as bulk Earth, while the composition inferred at 0446$-255$ lies in the region occupied by the primitive mantle and crust. When Na abundances are considered, two outliers are revealed in panel~(b): 0449$-259$ and 1350$-162$. They are accreting material greatly enhanced in Na and are located far from most comparison objects, in contrast to their proximity to the main body of comparisons in panel~(a). The apparent similarity of 1350$-162$ to crust material in panel~(b) is not replicated in panel~(a), highlighting the critical need to determine abundances for several elements. In panel~(c), the material at 0446$-255$ again lies close to the primitive mantle and crust, while the material at several other stars clusters near solar and bulk Earth abundances. The Ca depletion in the material polluting 2105$-820$ is not unprecedented among meteorites, but it stands out from the rest of the sample. Meanwhile 0449$-259$ lies away from the main locus of solar system objects, and would be even further removed if seen in the decreasing phase.

\subsection{Chi-squared analysis}
\label{subsectionChiSquaredResults}

The results presented in Fig.~\ref{figureElementRatios} allow a qualitative assessment of the sample, as set out in \ref{subsectionAbundanceRatios}. However, with several elements detected at each star there are many permutations of element ratios that could be considered. By contrast, the $\upchi^{2}$~procedure described in Section~\ref{subsectionChiSquaredComparison} allows all detected elements to be considered at once, and quantifies the similarity between the accreted material and solar system objects.

\begin{figure*}
 \includegraphics[width=\textwidth]{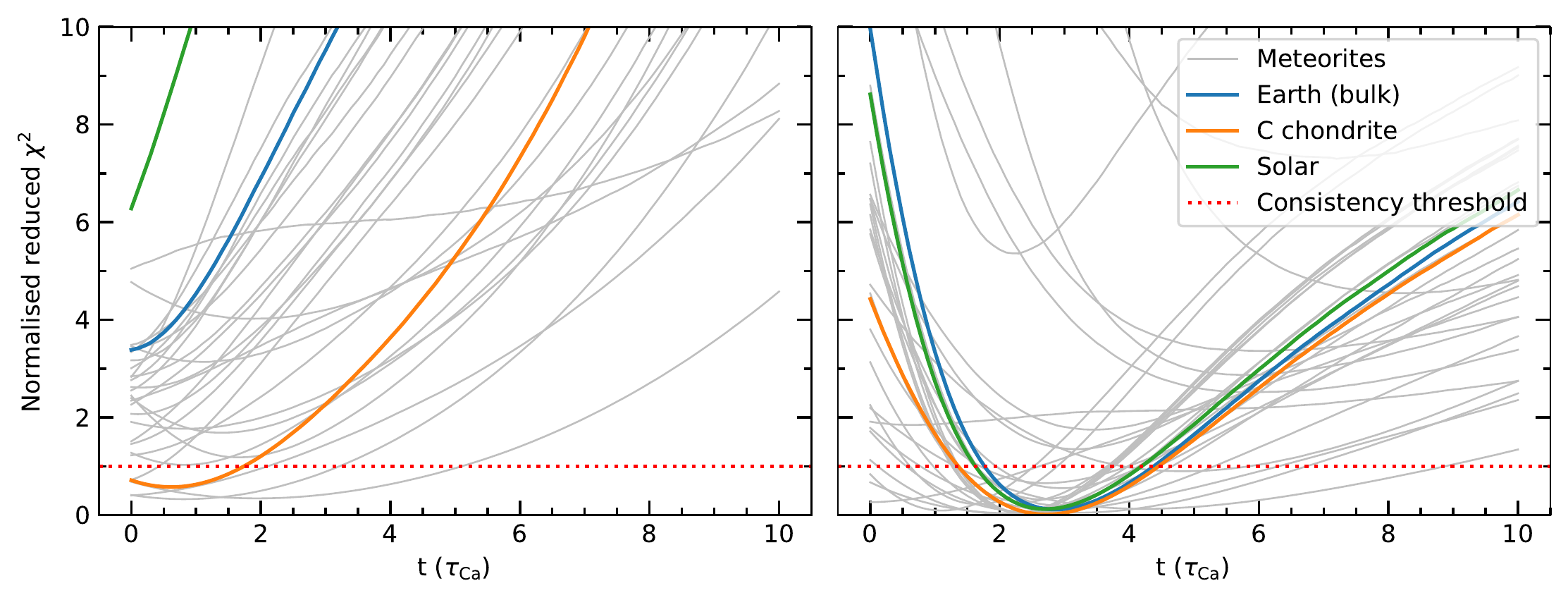}
 \caption{Reduced~$\upchi^{2}$ values at {0446$-255$}~(left) and {2216$-657$}~(right) for consistency between accreted and comparison materials in the decreasing phase. Accretion is in the increasing phase at $t=0$. The $\upchi^{2}$ values have been normalised to the 95~per~cent confidence level (the red dotted line), below which the test indicates consistency between the accreted and comparison materials.}
 \label{figureChiSquared}
\end{figure*}

\begin{figure*}
 \centering
 \includegraphics[width=\textwidth]{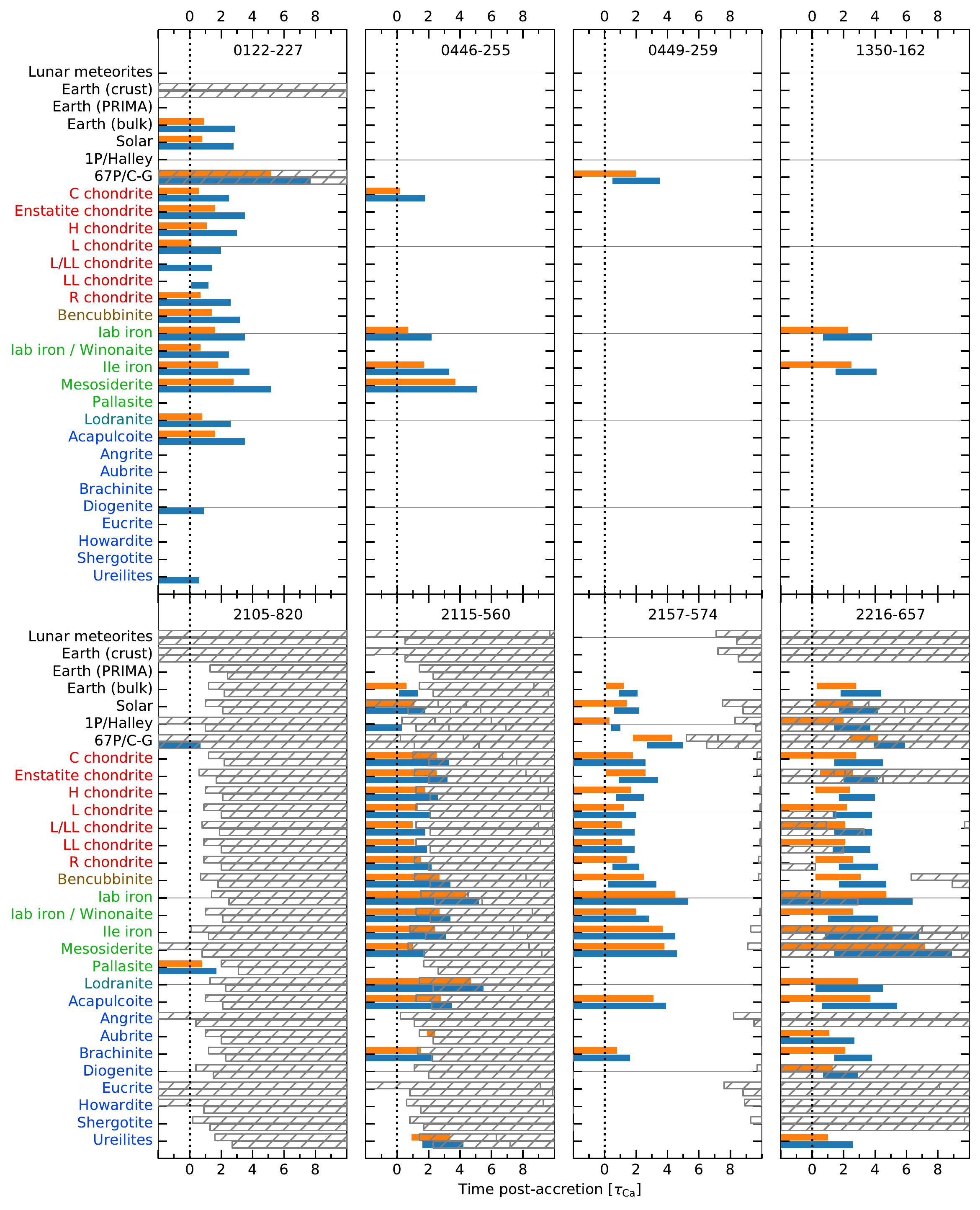}
 \caption{Results of the $\upchi^{2}$~analysis: coloured shading indicates consistency between accreted and comparison materials, with horizontal grid lines to guide the eye. Meteorite groups are ordered from chondrites~(red), through irons~(green), to achondrites~(blue). The region to the left of the dotted line at $t\leq0$ represents ongoing accretion, which ceases at $t=0$ and enters the decreasing phase for $t>0$. Results are given for each comparison assuming that accretion is in the increasing phase~(blue) or in steady state~(orange) at $t\leq0$. Grey hatching indicates where accretion of a particular composition would violate an observational upper limit. To aid interpretation of this Figure, an example is described in the text (Section~\ref{subsectionChiSquaredResults}.}
 \label{figureConsistency}
\end{figure*}

As the observed phase of accretion is unknown, all three phases are considered, via two calculations. First, for the increasing phase, the observed abundances are compared directly against solar system compositions, as in that phase the photospheric abundance ratios match those of the accreted material. Accretion is then assumed to switch off at time $t=0$, and the comparison is made against diffusion-modified abundance ratios of the accreted material (the solar system comparison). The calculation is performed across a grid of times in the decreasing phase up to $t=10$\,$\uptau_{\text{Ca}}$ after the accretion episode. Second, the photospheric abundances are modified to reflect accretion that has reached a steady state by the time accretion ceases at $t=0$, and the whole exercise is repeated. Results of the $\upchi^{2}$ analysis are shown in Figs.~\ref{figureChiSquared} and~\ref{figureConsistency}, as detailed below.

\begin{table*}
\begin{center}
\caption{Stellar parameters, photospheric log number abundances (relative to H or He), minimum $M_{\text{Z}}$ and $\dot{M}_{\text{Z}}$ in steady state.}
\label{tableResults}
\resizebox{\textwidth}{!}{ 
\begin{tabular}{lrrrrrrrrr}
\hline

WD									&0122$-$227		&0446$-$255		&0449$-$259		&1350$-$162		&2105$-$820		&2115$-$560		&2157$-$574		&2216$-$657		&2230$-$125$^{\rm a}$\\
\hline

\\
\multicolumn{10}{l}{\textit{Stellar Parameters:}}\medskip\\

Atmosphere									&He				&He				&He				&He				&H				&H				&H				&He				&H\\
$T_{\text{eff}}$ (K)						&$8380\pm170$	&$10\,120\pm200$	&$9850\pm160$	&$11\,640\pm290$	&$10\,890\pm380$	&$9600\pm340$	&$7010\pm250$	&$9190\pm210$	&$21\,170\pm740$\\
$\log{[g\,(\text{cm\,s}^{-2})]}$				&$8.06\pm0.04$	&$8.00\pm0.04$	&$8.04\pm0.03$	&$8.02\pm0.05$	&$8.41\pm0.08$	&$7.97\pm0.08$	&$8.06\pm0.08$	&$8.05\pm0.05$	&$7.93\pm0.08$\\
$M_{*}$\,($\text{M}_{\sun}$)						&$0.61$			&$0.58$			&$0.61$			&$0.60$			&$0.86$			&$0.58$			&$0.63$			&$0.61$			&$0.59$\\
$\log{(M_{\text{cvz}}/M_{*})}^{\rm b}$					&$-5.2$			&$-5.2$			&$-5.0$			&$-5.2$			&$-12.2$			&$-10.3$			&$-8.6$			&$-5.1$			&$-16.3$\\
$\log{[\uptau_{\text{Ca}}\,(\text{yr})]}$			&$6.2$			&$6.2$			&$6.3$			&$6.1$			&$0.6$			&$2.3$			&$3.6$			&$6.2$			&$-2.1$\\
$d$ (pc)								&64				&91				&67				&110			&16				&25				&29				&26				&120\\

\\
\multicolumn{10}{l}{\textit{Abundances:}}\medskip\\

H									&$-4.4$:\phantom{XXX}		&$-4.0\pm0.1$		&$-5.0\pm0.2$		&$-5.3\pm0.1$		&-\phantom{XXXX}				&-\phantom{XXXX}				&-\phantom{XXXX}				&$<-6.0$\phantom{XXX }			&-\phantom{XXXX}\\
C									&-\phantom{XXXX}				&-\phantom{XXXX}				&-\phantom{XXXX}				&-\phantom{XXXX}				&-\phantom{XXXX}				&$<-4.3$\phantom{XXX }			&$<-3.6$\phantom{XXX }			&-\phantom{XXXX}				&$-7.4$:\phantom{XXX}\\
N									&-\phantom{XXXX}				&-\phantom{XXXX}				&-\phantom{XXXX}				&-\phantom{XXXX}				&-\phantom{XXXX}				&$<-4.0$\phantom{XXX }			&$<-3.0$\phantom{XXX }			&-\phantom{XXXX}				&-\phantom{XXXX}\\
O									&$<-5.2$\phantom{XXX }			&$-5.8\pm0.1$		&$<-6.6$\phantom{XXX }			&$-6.2\pm0.1$		&-\phantom{XXXX}				&$<-5.0$\phantom{XXX }			&$<-3.8$\phantom{XXX }			&$<-6.5$\phantom{XXX }			&$<-6.0$\phantom{XXX }\\
Na									&-\phantom{XXXX}				&$-7.9\pm0.1$		&$-7.6\pm0.2$		&$-7.0\pm0.3$		&-\phantom{XXXX}				&-\phantom{XXXX}				&-\phantom{XXXX}				&$<-8.5$\phantom{XXX }			&-\phantom{XXXX}\\
Mg									&$-8.5\pm0.4$		&$-6.6\pm0.1$		&$-8.3\pm0.4$		&$-6.8\pm0.1$		&$-6.0\pm0.2$		&$-6.4\pm0.1$		&$-7.0\pm0.1$		&$-7.1\pm0.1$		&$<-6.4$\phantom{XXX }\\
Al									&-\phantom{XXXX}				&$-7.3\pm0.3$		&-\phantom{XXXX}				&-\phantom{XXXX}				&-\phantom{XXXX}				&$-7.6\pm0.1$		&$-8.1\pm0.1$		&-\phantom{XXXX}				&-\phantom{XXXX}\\
Si									&$<-7.6$\phantom{XXX }			&$-6.5\pm0.1$		&$<-7.3$\phantom{XXX }			&$-7.3\pm0.2$		&$<-5.5$\phantom{XXX }			&$-6.2\pm0.1$		&$-7.0\pm0.1$		&$<-7.0$\phantom{XXX }			&$-7.7\pm0.2$\\
P									&-\phantom{XXXX}				&-\phantom{XXXX}				&-\phantom{XXXX}				&-\phantom{XXXX}				&-\phantom{XXXX}				&-\phantom{XXXX}				&-\phantom{XXXX}				&-\phantom{XXXX}				&$<-8.0$\phantom{XXX }\\
Ca									&$-10.1\pm0.1$	&$-7.4\pm0.1$		&$-9.1\pm0.1$		&$-8.7\pm0.1$		&$-8.2\pm0.1$		&$-7.4\pm0.1$		&$-8.1\pm0.1$		&$-9.0\pm0.1$		&$-6.4$:\phantom{XXX}\\
Sc									&-\phantom{XXXX}				&$-10.9\pm0.2$	&-\phantom{XXXX}				&-\phantom{XXXX}				&-\phantom{XXXX}				&-\phantom{XXXX}				&-\phantom{XXXX}				&-\phantom{XXXX}				&-\phantom{XXXX}\\
Ti									&-\phantom{XXXX}				&$-8.8\pm0.1$		&$-10.7\pm0.2$	&$-10.0\pm0.1$	&-\phantom{XXXX}				&-\phantom{XXXX}				&-\phantom{XXXX}				&$-10.6\pm0.1$	&-\phantom{XXXX}\\
V									&-\phantom{XXXX}				&$-10.0\pm0.3$	&-\phantom{XXXX}				&-\phantom{XXXX}				&-\phantom{XXXX}				&-\phantom{XXXX}				&-\phantom{XXXX}				&-\phantom{XXXX}				&-\phantom{XXXX}\\
Cr									&-\phantom{XXXX}				&$-8.5\pm0.1$		&-\phantom{XXXX}				&$-9.0\pm0.2$		&-\phantom{XXXX}				&-\phantom{XXXX}				&-\phantom{XXXX}				&-\phantom{XXXX}				&-\phantom{XXXX}\\
Mn									&-\phantom{XXXX}				&$-9.1\pm0.1$		&-\phantom{XXXX}				&-\phantom{XXXX}				&-\phantom{XXXX}				&-\phantom{XXXX}				&-\phantom{XXXX}				&-\phantom{XXXX}				&-\phantom{XXXX}\\
Fe									&$-8.5\pm0.2$		&$-6.9\pm0.1$		&$-7.9\pm0.2$		&$-7.1\pm0.1$	&$-6.0\pm0.2$		&$-6.4\pm0.1$		&$-7.3\pm0.1$		&$-8.0\pm0.2$		&$<-6.0$\phantom{XXX }\\
Ni									&-\phantom{XXXX}				&$-8.2\pm0.1$		&$-8.4\pm0.2$		&-\phantom{XXXX}				&-\phantom{XXXX}				&-\phantom{XXXX}				&$-8.8\pm0.1$		&-\phantom{XXXX}				&$<-7.3$\phantom{XXX }\\
Sr									&-\phantom{XXXX}				&$-11.1\pm0.2$	&-\phantom{XXXX}				&-\phantom{XXXX}				&-\phantom{XXXX}				&-\phantom{XXXX}				&-\phantom{XXXX}				&-\phantom{XXXX}				&-\phantom{XXXX}\\

\\
\multicolumn{10}{l}{\textit{Heavy element masses and accretion rates:}}\medskip\\

$\log{[\dot{M}_{\text{Z}}\,(\text{g\,s}^{-1})]}$	&$7.0\pm0.2$ &$9.1\pm0.1$ &$7.8\pm0.1$ &$8.8\pm0.1$ &$8.8\pm0.2$ &$8.8\pm0.2$ &$8.5\pm0.1$ &$7.8\pm0.1$ &$6.7\pm0.1$\\
$\log{[M_{\text{Z}}\,(\text{g})]}$				&$20.8\pm0.8$ &$23.0\pm0.7$ &$21.7\pm0.7$ &$22.6 \pm 0.6$ &$16.9 \pm 0.7$ &$18.5 \pm 0.5$ &$19.4 \pm 0.3$ &$21.8 \pm 0.7$ &$12.0 \pm 0.2$\\

\hline

\end{tabular}} 
\end{center}

\flushleft
{\footnotesize {Poorly-constrained quantities are indicated with a colon.

$^{\rm a}$Reported abundances assume photospheric origin, but C and Ca are likely contaminated by interstellar absorption; see Section~\ref{sectionResults}.

$^{\rm b}$All stars here have a surface convection zone except 2230$-$125, where the mixing layer is defined as the region above optical depth $\uptau_{\text{R}}\leq5$.}}

\end{table*}

In general, inferred abundances become consistent with fewer comparisons as accretion progresses further into the decreasing phase. Fig.~\ref{figureChiSquared} shows the $\upchi^{2}$ values for 0446{$-255$} (a typical case) and 2216{$-657$} (an exception), for accretion in the increasing phase that then ceases and enters the decreasing phase. Only a few comparisons are shown in colour, with the remainder in grey to maintain readability. The horizontal dotted line shows the 95~per~cent confidence level; a $\upchi^{2}$ value below this level indicates consistency between the observed abundance and the solar system comparison. The key feature of these plots is the shape of the curves. At 0446$-255$, the $\upchi^{2}$ values tend to increase along the x-axis, moving away from consistency as soon as accretion ceases. This suggests that accretion is ongoing. By contrast, at 2216$-657$ the $\upchi^{2}$ values decrease initially, reach their minima around 2.7\,$\uptau_{\text{Ca}}\approx5$\,Myr, and then increase. Assuming the accreted material is typical of rock, this is a clear indication that the star is observed in the decreasing phase. Applying this analysis to the rest of the sample, there is no strong evidence against ongoing accretion at any star other than 2216$-657$. For 0122$-227$, the $\upchi^{2}$~values change only gradually throughout the decreasing phase, while for 0449{$-259$} and 1350$-162$ interpretation is complicated by the unusual Na abundances discussed below. Accretion should be in a steady state at the H-atmosphere stars due to their short sinking time-scales, and the $\upchi^{2}$ values are notably consistent with this expectation.

Fig.~\ref{figureConsistency} shows the results of the $\upchi^{2}$~comparisons between all stars in the sample and all comparison objects, omitting 2230$-125$ as it has only one secure photospheric abundance measurement. Coloured shading shows where the $\upchi^{2}$ values indicate consistency, which corresponds to the segment of the curve below the 95~per~cent confidence level in Fig.~\ref{figureChiSquared}. As described above, the analysis has been performed for accretion in both increasing and steady-state phases, then switching to the decreasing phase. Blue shading indicates consistency during or after the increasing phase, and orange shading indicates consistency during or after the steady-state phase. Ongoing accretion is represented by the region left of the dotted line at $t\leq0$, where accretion switches off for $t>0$. In some cases, consistency is indicated by the $\upchi^{2}$ results (which depend only on the detected elements), while contra-indicated by abundance upper limits, because observable quantities of those elements would have been present for that composition. This is particularly true in the decreasing phase. Violation of one or more abundance upper limits is indicated in Fig.~\ref{figureConsistency} by grey hatching. The expected abundances for a given composition are taken relative to Mg for consistency, but this test is not strict as Mg abundances will vary between stars in the sample. Nevertheless, it is clear that abundance upper limits can aid interpretation. For example, three stars show consistency with ongoing accretion of material like the dust of comet \mbox{67P/C--G}, but only 0449$-$259 is compatible with the upper limits.

As Figure~\ref{figureConsistency} is complex, it is helpful to illustrate its interpretation by way of an example. Reading downwards from the top, the results for 0122$-227$ show that accretion of material sharing the composition of lunar meteorites would not produce photospheric abundances consistent with those observed, regardless of the phase of accretion. The same is true for the crust of the Earth, where the grey hatching indicates that one or more elements in the accreted material would be present in quantities that violate an observational upper limit. Skipping down to the first coloured bars, these show that the observed abundances are consistent with the accreted material having bulk Earth composition. The blue bar shows consistency if increasing-phase accretion is ongoing, or if it has ceased and up to 2.8\,$\uptau_{\text{Ca}}$ have elapsed in the decreasing phase. Likewise, the orange bar shows consistency with ongoing accretion in steady-state, or up to 0.8\,$\uptau_{\text{Ca}}$ into the decreasing phase. Skipping down again, coloured bars show that the dust of comet \mbox{67P/C--G} could have been consistent with the accreted abundances, were it not for an observational upper limit violation across all phases of accretion, as indicated by the grey hatching. Finally, a bit further down, comparison with the LL~chondrite composition shows consistency only during a small range in the decreasing phase, and only if accretion ceased while still in the increasing phase; the observed abundances are \textit{not} consistent with ongoing accretion in the increasing phase, as indicated by the absence of colour to the left of the dotted line at $t\leq0$.

The results show that the accreted abundances inferred at each of the stars are consistent with at least one solar system composition. Several stars are accreting material consistent with numerous comparison objects, most likely due to similarity of the accreted material to rock, and the wide spread of abundance measurements within meteorite classes. Two of the H-dominated stars have similar results, showing consistency mostly with chondritic and iron meteorites, bulk Earth, and solar abundances. However, some stars are consistent with far fewer comparisons, tending towards iron meteorites. In particular, 0449$-$259, 1350$-$162, and 2105$-$820 have unusual abundances and show consistency with only one or two comparisons. Note that 0122$-227$ has few detected elements, with larger uncertainties than the other stars, so its $\upchi^{2}$ results are less constraining.

\subsection{Individual objects}
\label{subsectionIndividualObjects}

\textit{0122$-$227} -- Previously classified as spectral type DZ \citep{Friedrich2000}, the spectrum now reveals itself as DZA. The H\,$\upalpha$~line is sharper and deeper in the data than in the model, and thus the H abundance determination is uncertain (\ref{tableResults}). The H\,$\upalpha$ line velocity matches that of the photospheric metals, and because Balmer absorption is not observed in the ISM, this feature must be intrinsic to the star. With only three elements measured, little can be said about the composition of the accreted material, beyond that it is broadly consistent with solar system objects, including bulk Earth.

\medskip
\textit{0446$-$255} -- The $\upchi^{2}$ results suggest ongoing accretion, as shown in Fig.~\ref{figureChiSquared}. This is supported by the element ratios shown in Fig.~\ref{figureElementRatios}, where the inferred abundances depart from those typical of meteorites once accretion ceases. While the accreted material has broadly similar abundance ratios to solar system materials, it tends more toward to crust or mantle compositions than bulk Earth. The $\upchi^{2}$ analysis is consistent with some iron-rich meteorites, or the primitive C~chondrites (although Sc, V, and Sr are not measured for many meteorite groups, and thus are not considered).

Detailed comparison with Earth abundances shows that the accreted material is a poor match for the crust, with some elements (particularly Cr, Si, and Sr) deviating by up to 2~dex compared to the major rock-forming elements. The other components of the Earth are better matched, but not consistent. Compared to bulk Earth, the accreted material is depleted in Fe and Ni (relative to Si or Mg) and enhanced in Na, Ca, and Ti, while compared to the primitive mantle, it is depleted in Mg and enhanced in Fe and Ni. However, by considering a mixture of the primitive mantle, core, and crust of the Earth, a closer match is found. Using logit-transformed proportions for the abundances (Eqn.~\ref{equationLogit}), $\upchi^{2}$~minimization finds crust\,:\,mantle\,:\,core ratios of 0\,:\,96\,:\,4 in the increasing phase and 9\,:\,80\,:\,11 in the steady state. A composition of mantle rock mixed with a small amount of core material is consistent with the depletion of Fe compared to bulk Earth, and enhancement of Ni compared to the primitive mantle. The ratio $\ratio{Fe}{Ni}=21\pm7$ is consistent with that of the bulk Earth, but not the primitive mantle. Such a composition, the $\ratio{Fe}{Ni}$ ratio, and the minimum $M_{\text{Z}}=10^{23}$\,g, are all consistent with accretion of a minor planet with a small core.

The O budget can be examined by considering how much could be delivered in the form of metal oxides \citep{Klein2010}. The numerical abundance of O shows an excess of $10\pm15$~per~cent in the increasing phase, and a deficit of $19\pm18$ in steady-state, i.e. it is consistent with accretion of anhydrous rock. In the extreme case of treating all observed Fe and Ni as free metal, as in a planetesimal core, these values become 18~per~cent (excess) and 4~per~cent (deficit), respectively. Therefore, within the uncertainties, the presence of some core material is compatible with the O budget.

\medskip
\textit{0449$-$259} -- The $\ratio{Ni}{Fe}$ ratio is enhanced by a factor of six compared to bulk Earth, while $\ratio{Na}{Fe}$ is enhanced by a factor of 60. Neither ratio is typical of solar system objects. If seen in steady-state, Na accounts for 20~per~cent by mass of the accreted material, though this falls if seen in the decreasing phase, since Na has the second-longest sinking time of the elements observed. The $\ratio{Ca}{Fe}$ and $\ratio{Mg}{Fe}$ ratios (Fig.~\ref{figureElementRatios}) suggest that the enhancement in Na is real, rather than being a side-effect of Fe depletion. Indeed, $\ratio{Na}{Ca}=8\pm3$ is higher than previously reported at any polluted white dwarf (\citealt{Blouin2019}; the photospheric ratio found in that study is $\ratio{Na}{Ca}=6\pm3$, compared to $19\pm8$ for 0449$-$259).

Taken at face value, the median of the $\upchi^{2}$~minima across all comparisons suggests that 0449$-259$ is observed well into the decreasing phase -- around three sinking times if accretion had reached steady-state before it ceased, and around five sinking times otherwise. However, this is due to the the unusual Na abundance: if it were excluded, the $\upchi^{2}$~minima would indicate ongoing accretion. Also, Mg is under-abundant compared to solar system objects, whereas it would be expected to become enhanced in the decreasing phase, having an even longer sinking time-scale than Na. Moreover, the abundances are consistent with \mbox{67P/C--G} if seen in steady state (with the caveat that abundances of Ti and Ni are not available for the comet, and so are not considered in the $\upchi^{2}$ analysis). Therefore, it appears more likely that the unusual Na abundance at this star is shared by the accreted material, rather than being due to observation in the decreasing phase.

\medskip
\textit{1350$-$162} -- The photospheric abundances would be a good match for both solar and bulk Earth, were Na not over-abundant by more than a factor of 100 compared to Ca or Fe, with $\ratio{Na}{Ca}=11\pm9$. That quantity is less certain than at 0449$-259$, as only a single line of the Na~doublet is visible in the spectrum, but nevertheless it is clearly enhanced compared to solar system values. As with {0449$-259$}, the $\upchi^{2}$~minima suggest that accretion may be in the decreasing phase, if it never reached a steady state. However, unlike at 0449$-259$, this is not an artefact of the unusual Na abundance: for accretion that ceased during the increasing phase, the median of $\upchi^{2}$~minima moves from 1.8\,$\uptau_{\text{Ca}}$ with Na, to 1.6\,$\uptau_{\text{Ca}}$ without Na. The exponential divergence of abundances in the decreasing phase would only reduce the inferred $\ratio{Na}{Ca}$ ratio for the accreted material by a factor of two on that time-scale, so the phase of accretion is not tightly constrained by abundance ratios. The increasing phase appears the least likely, as neither of the comparison compositions that show consistency with the measured abundances do so in the increasing phase (Fig.~\ref{figureConsistency}).

The O budget is not helpful in further constraining the phase that accretion is seen in. For accretion of anhydrous rock that has reached steady state, the budget balances within the uncertainties for ongoing accretion and up to 1.5\,$\uptau_{\text{Ca}}$ thereafter. For accretion in the increasing phase, there is a marginal O excess for ongoing accretion but the budget balances at 1--3\,$\uptau_{\text{Ca}}$ into the decreasing phase.

\medskip
\textit{2105$-$820} -- The H\,$\upalpha$ line does not have a sharp core, being instead slightly flattened. This is likely due to Zeeman splitting of the line in this known magnetic star (field strength $B\approx43$\,kG), the hottest of that class to show pollution \citep{Koester1998,Landstreet2012, Kawka2019}. The material polluting this star is strikingly deficient in Ca compared to other polluted white dwarfs \citep{Kawka2014}. When compared to a recent study of 230~such stars \citep{Hollands2018analysis}, the ratios $\ratio{Fe}{Ca}=200\pm100$ and $\ratio{Mg}{Ca}=110\pm50$ would be outliers beyond $3\upsigma$, each comparable only to one or two other stars among that sample. Only 1011$+570$ has similar ratios between Ca, Fe, and Mg, albeit with larger uncertainties \citep{Wolff2002}. Both the $\ratio{Ca}{Fe}$ and $\ratio{Ca}{Mg}$ ratios are an order of magnitude below those of bulk Earth, and two below those of its crust. While only an upper limit can be placed on the Si abundance, it is stringent enough to be useful. The upper limits on the $\ratio{Si}{Fe}$ and $\ratio{Si}{Mg}$ ratios lie between the values for bulk Earth and its crust. Thus, it appears that the accreted material bears little similarity to a planetary crust.

The severe Ca depletion could indicate a high olivine content. The only consistent comparison is the pallasite meteorite group, whose members consist of olivine crystals embedded in a $\ratio{Fe}{Ni}$ matrix. Two similar meteorite groups have a $\upchi^{2}$~value just outside of consistency: the IAB irons and the lodranites. A strict interpretation of the $\upchi^{2}$~test would ignore these meteorite groups, but this star only has three elements available for the comparison, and the test as used here is indicative but not conclusive. All three comparisons have similar origins. Pallasites are thought to have been formed at the core-mantle boundary of a differentiated body, and extracted later in a collision \citep{Yang2010}. The Ca-poor lodranites are thought to originate in a partially-melted layer a few km beneath the surface of a differentiated body, and to have been excavated by an impact \citep{Henke2014,Neumann2018}. The IAB meteorites appear to have a more complex history, sampling a partially-differentiated body that was disrupted while the core was molten and then reassembled \citep{Hunt2018}. Consistency with such comparisons points to this star having accreted material that originated within a differentiated object.

While not considered here, previous studies of this star have found a radiative atmosphere to provide a better fit than a convective one \citep{Bedard2017, GentileFusillo2018_2105}. Convection may be suppressed by a magnetic field, particularly at the surface \citep{Tremblay2015}. If the photospheric metals are not mixed in a convection zone, that could imply a higher $\dot{M}_{\text{Z}}$ than inferred here, where it is already near the top of the range observed at H-dominated stars. However, this is not explored here, as atmospheric models do not yet account for the effects of a magnetic field on sinking time-scales. It is noteworthy that the mass determined here is significantly higher than the average for white dwarfs of $0.6\,\text{M}_{\sun}$ \citep{Koester2015,Kepler2016}. There is no consensus on the origin of white dwarf magnetic fields, but one hypothesis for $B\gtrsim1$\,MG is a binary merger \citep{Tout2008}, which would be consistent with the high mass. If the white dwarf evolved as a single star, the progenitor mass would be typical of a late B-type star, a spectral class whose planetary systems are not yet well characterised.

\medskip
\textit{2115$-$560} -- The $\upchi^{2}$~analysis shows that the accreted material is consistent with solar abundances, bulk Earth and many meteorites, especially chondrites. The widest deviation from solar ratios is the slightly elevated $\ratio{Si}{Al}=25\pm8$. However, this ratio is well within the observed spread in solar system meteorites. This material therefore appears to be primitive, i.e. unprocessed condensates from a cloud similar to the solar nebula. The upper limit on the O abundance leads to a limit on the ratio $\ratio{O}{Ca}<70$ that is comparable to the solar value, suggesting that the accreted material does not have a large water ice component.

\medskip
\textit{2157$-$574} -- A more careful approach to the modelling is required here than for the other stars, as the atmosphere is sufficiently cool that H is mostly neutral. Thus, most free electrons originate from metals, and some observed metal lines are sufficiently strong to influence the fitting of stellar parameters. In order to find the overall best fit, abundances are determined for multiple atmospheres, using a grid of stellar parameters centred on those indicated for a pure H atmosphere.

The abundances are for the most part a close match for bulk Earth, with only Ni showing a small departure. The $\ratio{Fe}{Ni}$ ratio is $29\pm11$, the highest seen amongst this sample of stars. However, it is closer to the value of 18 for bulk Earth than the crust value of 840. A recent study of 38~solar analogues found $\ratio{Fe}{Ni}$ ratios within the range 11--29 \citep{Lopez-Valdivia2017}. Thus, the ratio in the material polluting this star does not appear unusual, assuming that the solar analogues can be used as a proxy for its younger and more massive main-sequence progenitor.

\medskip
\textit{2216$-$657} -- This star is likely observed in the decreasing phase. If accretion is ongoing, then the abundances at this star are challenging to interpret, since Mg accounts for over 70~per~cent of the observed $M_{\text{Z}}$ by mass. However, Mg sinks around 2.5~times more slowly than the other detected elements, and it will therefore come to dominate in the decreasing phase. This is the only star here with $\ratio{Mg}{Fe}$ higher than that of bulk Earth, approaching it as time increases in the decreasing phase, while the other stars start at or below that ratio and depart further over time (Fig.~\ref{figureElementRatios}). As shown in Section~\ref{subsectionChiSquaredResults}, the abundances become most consistent with solar system objects if accretion ceased around 5~Myr (2.7\,$\uptau_{\text{Ca}}$) ago if accretion had not reached a steady state, or around half that time if it had. A broad range of materials are consistent at that time. Since Na sinks at a comparable rate to Mg it would also appear enhanced. However, the upper limit determined here is only stringent enough to rule out a composition similar to crust material, even in the decreasing phase (Fig.~\ref{figureElementRatios}). Both $M_{\text{Z}}$ and the relatively low, time-averaged $\dot{M}_{\text{Z}}$ are compatible with accretion having ceased. In the case that accretion ended 5\,Myr prior in this star, the minimum value of $M_{\text{Z}}$ would increase by an order of magnitude, placing it at the high end of the range in this sample.

Lines of C and Si have been detected in ultraviolet spectra of this star \citep{Wolff2002}. The data are of modest quality, and were analysed using different atmospheric models, but if combined with the data presented here would imply that the photospheric metals are spectacularly rich in C. The sinking time-scale of C is longer than those of the detected elements, suggesting that like Mg it has become concentrated in the decreasing phase. For ongoing accretion, C would account for tens of per~cent of the total mass, two orders of magnitude more than in bulk Earth. Such an abundance may be typical of comets, but the upper limits on H, O, and Na suggest that the accreted material is not rich in volatiles, and that the decreasing phase is the more likely cause of the C enhancement. Similar conclusions could be reached using the abundances presented in a recent analysis of this star \citep{Allard2018}, where measurements in common with this study agree within the uncertainties.

A sanity check was made against the conclusion that this star is seen in the decreasing phase, using randomly-generated data. For each combination of detected elements, 200 sets of abundances were drawn from a uniform distribution over a range of 3.0\,dex with 0.1\,dex uncertainties, and stellar temperatures were drawn from a uniform distribution over the range 6000--25\,000\,K. When only Ca, Mg, and Fe are considered, half of the sets find a match with one or more comparisons, split equally between ongoing accretion and the decreasing phase. However, fewer matches are found as more elements are used, and those matches are weighted towards ongoing accretion. For the elements seen at {2216$-657$}, only a few per~cent of the random sets produce matches in the decreasing phase. Thus, it appears that material with unusual chemistry would not typically masquerade as rocky material seen post accretion, provided that several elements are measured. Thus, all the available evidence points to the atmospheric metals being the remains of rocky material accreted a few sinking time-scales ago.

\medskip
\textit{2230$-$125} -- The atmosphere of 2230$-125$ has not yet cooled sufficiently for convection. Abundances are determined assuming that the material responsible for forming the lines is mixed to an optical depth $\uptau_{\text{R}}=5$. The stellar parameters are well determined from the optical spectra alone, being consistent with the result obtained by also considering the \textit{Gaia} parallax, photometry, and the ultraviolet spectrum.

Lines due to C, N, Na, Si, S, and Ca are present in the optical and ultraviolet spectra. Unfortunately, interstellar absorption features prevent detailed analysis of potential photospheric metal lines. For example, \ion{Na}{I}~D is clearly of interstellar origin. Both components of the doublet are shifted by +7\,km\,s\textsuperscript{-1} relative to the Balmer lines. The same is true for the \ion{Ca}{II}~H and K lines, whereas a \ion{Si}{II} line at {1260\,\AA} is shifted by $-9$\,km\,s\textsuperscript{-1} relative to the Balmer lines. Lines are present at {1265\,\AA} due to \ion{Si}{II}, and at {1298\,\AA} due to \ion{Si}{III}, where both lines have the expected velocity of the photosphere. Because these transitions should not be excited in the ISM, they represent the only secure detections of photospheric metals in this star.

The abundance ratios strengthen the case for interstellar line contamination: even if the upper limits for Fe and Mg were to be treated as detections, the $\ratio{Ca}{Fe}$ and $\ratio{Ca}{Mg}$ ratios would both be near the upper bounds of the ranges observed in solar system objects. Meanwhile, the $\ratio{Ca}{Si}$ ratio is 100~times higher than any known solar system comparison and 10~times higher than observed in any polluted white dwarf \citep{Gansicke2012}. The $\ratio{C}{Si}$ ratio is at the top end of the solar system range, but is not so high as to rule out purely photospheric absorption by C. With Si being the only element where an abundance is securely determined, the chemistry of the accreted material remains unconstrained.

\section{Discussion}
\label{sectionDiscussion}

While confirming the apparent near-ubiquity of volatile-poor, rocky material polluting white dwarfs, this sample demonstrates considerable diversity. Compositions are consistent with a range from primitive, through solar or bulk Earth analogues, to parts of differentiated bodies. No excesses of O are found that would suggest the accretion of water-rich material, and no other volatiles are detected. In the broadest sense there are no surprises, but a closer look reveals interesting details, such as the exceptional Ca deficiency at 2105$-820$, and the Na enhancements at 0449$-259$ and 1350$-162$. While only 2115$-560$ shows an infrared excess consistent with circumstellar dust \citep{vonHippel2007,Mullally2007}, it nevertheless appears likely that most stars in the sample are presently accreting material. The exception is 2216$-657$, showing clear signatures of a system observed in the decreasing phase.

This sample adds to the list of stars thought to be accreting material originating in differentiated bodies, suggesting that the violent events that liberate such material are common in exoplanetary systems. These evolved stars therefore provide a line of evidence for planetesimal interactions independent and complementary to debris discs observed in main sequence systems that are thought to arise from such interactions, such as $\upbeta$~Pictoris \citep{Okamoto2004}. As detailed in Section~\ref{subsectionIndividualObjects}, the material accreted by 0446$-255$ is a good match for a minor planet with a small Fe--Ni core. However, Ca and Mg appear somewhat enhanced and depleted, respectively, which if seen in isolation might indicate a crust-like composition. Thus, a different conclusion might have been reached for this star, had fewer elements been available. This underscores the value of high S/N spectra of polluted stars, so that unusual abundances of one or two elements do not confuse the interpretation.

\subsection{Sodium}
\label{subsectionDiscussionSodium}

A wide range of Na abundances are seen in the sample. Potential causes include scatter in stellar chemistry, contamination by interstellar lines, formation conditions in the protoplanetary nebula, and subsequent radial transport. These are considered below, with reference to stars from the literature \citep{Zuckerman2007,Zuckerman2011, Klein2010,Klein2011, Farihi2011,Farihi2013, Koester2011,Xu2014,Raddi2015,Hollands2017abundances,Blouin2019}.

\begin{figure}
 \includegraphics[width=\columnwidth]{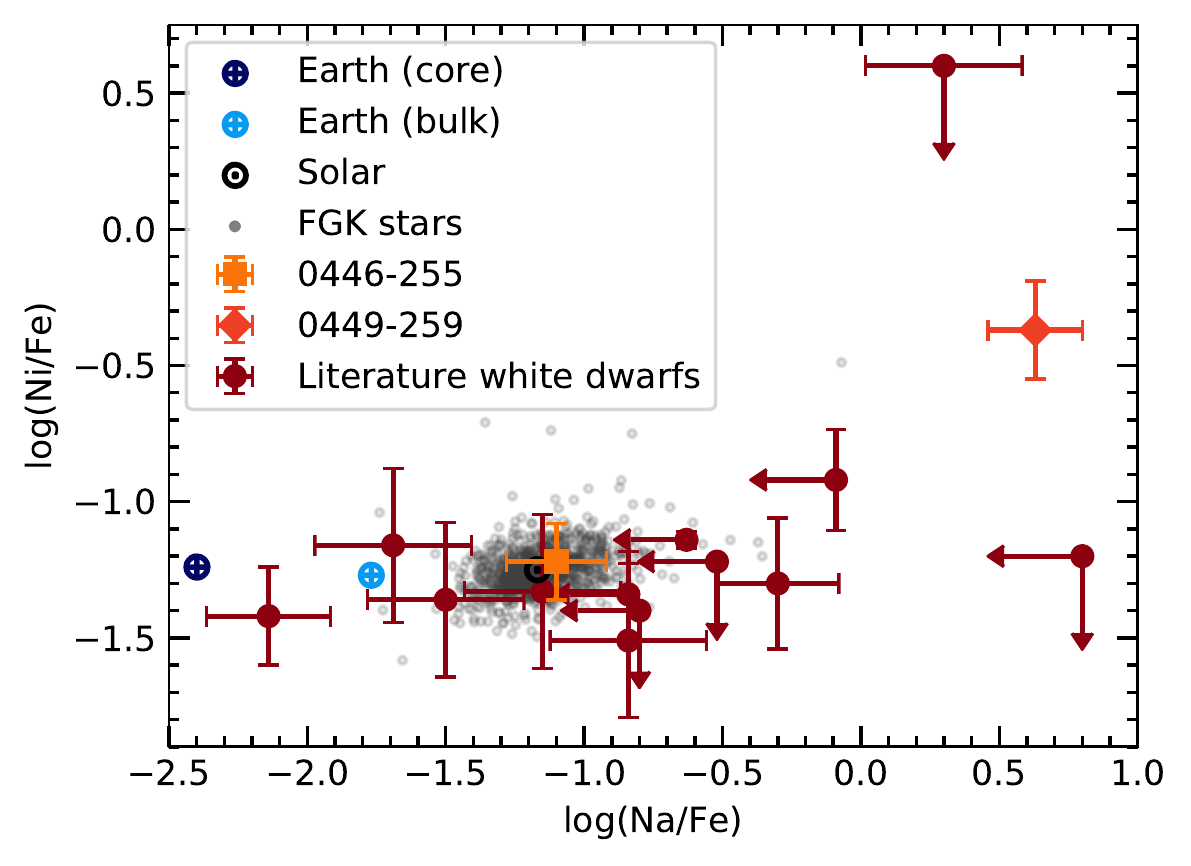}
 \caption{Photospheric log number abundance ratios of Ni and Na against Fe.}
 \label{figureNiNaFe}
\end{figure}

There is a correlation between $\ratio{Na}{Fe}$ and $\ratio{Ni}{Fe}$ ratios in solar-metallicity stars, and similar relations have been found in multiple lower-metallicity populations such as the thick disc, the halo, and in dwarf spheroidal galaxies \citep{Nissen2015}. Few polluted white dwarfs have reported measurements for both elements, though some upper limits are available. Fig.~\ref{figureNiNaFe} shows abundances and limits for the stars in this study, white dwarfs from the literature, and 1000 local FGK stars \citep{Hinkel2014}. No trend can be discerned for the polluted stars where both elements are measured. Including upper limits as if they are measurements introduces a correlation, however this is likely to reflect the limits having been imposed in a correlated fashion by the quality of the data. The FGK~stars show a correlation, but across a much smaller range than that covered by polluted white dwarfs. It is interesting that both Na and Ni appear enhanced at {0449$-259$} compared to other rock-forming elements, but overall the spread of Na abundances in exoplanetary material does not appear to be related to stellar chemistry.

The \ion{Na}{i}\,D doublet is a prominent line in the ISM \citep{Murga2015} and thus interstellar contamination is a potential concern for any star, as encountered for {2230$-125$}. While the Na lines in three additional targets (0446$-255$, 0449$-259$, 1350$-162$) might be a blend of photospheric and interstellar components, there is no evidence of line blending in the $R\approx9000$ spectra. Furthermore, it would be highly improbable that the line velocities match in all three systems. Most stars in this sample lie within 100\,pc and the Local Bubble where little interstellar material is expected \citep{Redfield2006}, but it is noteworthy that {2230$-125$} is just outside this radius at 125\,pc and appears to have interstellar absorption lines.

Given that neither interstellar material nor progenitor star chemistry appear to be major factors in determining Na abundance, intrinsic sources are considered. The moderate volatilility of Na may be relevant, as it is more sensitive to local conditions in the protoplanetary disc than refractory elements, such as Ca or Al. Thus, the wide range in abundances may arise from volatile depletion in the inner regions of the host system, during the formation of the bodies whose remnants are now being accreted, and simulations appear to support this \citep{Harrison2018}. Equally, subsequent radial transport may lead to enhancement of volatiles in bodies in the inner regions. The thermal history of the accreted bodies may also be important, especially for rubble-pile asteroids with large surface-area-to-volume ratios, as they will experience significant heating during the giant branch and planetary nebula phases of the host star \citep{Veras2016}. Simulations that further explore these scenarios may provide a context in which Na abundances can be interpreted.

Another potential source of Na enhancement is cryovolcanism. The \textit{faculae} (bright spots) on 1\,Ceres have been interpreted as concentrations of Na\textsubscript{2}CO\textsubscript{3} and other salts, brought to the surface in liquid water that sublimated \citep{DeSanctis2016}. However, this possibility is at odds with the overall abundances at 0449$-259$ and 1350$-162$ that do not otherwise suggest that fragments of crust material are being accreted.

The final hypothesis considered is that anomalous Na abundances are the last traces of past accretion events, whose signatures have since been diluted by more recent accretion. The long sinking time-scale of Na relative to other elements, and the consistency with some comparisons when it is removed from the analyses here, would support this. Such past accretion events could have involved crust-like material: the $\ratio{Na}{Mg}$ ratio is 50 times higher in the crust than in the bulk Earth, and so there would be no corresponding enhancement in slow-sinking Mg during the decreasing phase. However, a proper assessment of this scenario would require modelling that is beyond the scope of this study.

There may be similarities between the Na-rich stars in this study and 2354$-211$, a cool DZ star with a strong Na absorption feature \citep{Blouin2019}. The authors of that study note that the photospheric abundances of 2354$-211$ appear consistent with the dust at comet \mbox{67P/C--G}. The same is found here at 0449$-$259, albeit for steady-state accretion. There is no strong evidence that either star is in the decreasing phase: 2354$-211$ has unremarkable ratios between Fe, Ca, and Mg. It is also noted that comets are orders of magnitude too small to account for the observed $M_{\text{Z}}$ at 2354$-211$, which is again the case for 0449$-$259. These stars may be members of the same class, polluted by material with an origin unlike that found in the solar system.

\subsection{The decreasing phase}
\label{subsectionDiscussionPostAccretion}
The exponential divergence of elemental abundances once accretion ceases can lead to unusual ratios, such as those seen at 2216$-657$. The fact that those abundances evolve towards a rocky composition when traced back in time, but a random set of abundances do not, strongly suggests that observation post-accretion is the correct interpretation. If further examples of decreasing phase systems can be identified, they may allow insights into the accretion process.

Below around 20\,000\,K, the sinking time for Mg is longer than that of either Ca or Fe, which are similar to each other. Thus, generally the $\ratio{Mg}{Fe}$ ratio will grow larger over time, while the $\ratio{Ca}{Fe}$ ratio will remain approximately constant. Lines of these three elements are seen at many polluted stars, and thus a preliminary diagnostic presents itself: assuming the $\ratio{Ca}{Fe}$ ratio does not indicate that the material was deficient in Fe, a high $\ratio{Mg}{Fe}$ ratio is an indicator that the system may be in the decreasing phase. The $\upchi^{2}$ minima can provide a stronger indication of the state of accretion than the $\ratio{Mg}{Fe}$ ratio. The sanity check described in Section~\ref{subsectionIndividualObjects} shows the more elements that are available, the more reliable the $\upchi^{2}$ minima method becomes as a diagnostic. However, as seen in Section~\ref{sectionResults}, results can be inconclusive for stars displaying unusual element ratios, especially where few elements have been detected. Thus, for a reliable identification of a post-accretion system, other information should also be considered, such as $M_{\text{Z}}$ as a function of time.

Sinking times of photospheric metal are the same within a factor of a few. Thus, a given inferred $M_{\text{Z}}$ increases by a factor of order $e$ per sinking time after accretion switches off. If seen around 10 sinking times post-accretion, the abundances in this study would imply such high values of $M_{\text{Z}}$ during accretion that the steady-state $\dot{M}_{\text{Z}}$ required to support them would enter the regime of novae ($\dot{M}\gtrsim10^{13}$\,g\,s\textsuperscript{-1}; \citealt{Yaron2005}). Therefore, a system whose abundances suggest that it is observed more than a few sinking times post-accretion would be statistically unlikely. The frequency of polluted stars in the decreasing phase will depend on the accretion process. For example, abundance ratios retain their steady-state values if the accretion rate decays exponentially \citep{Koester2009}. Therefore, that model is ruled out at any star observed where abundances have have diverged in the decreasing phase, such as 2216$-657$.

Accretion at 2216$-657$ has been of an idiosyncratic nature, in that the photospheric abundances appear dominated by a single event. This is interesting in the context of planetesimal accretion models that aim to reproduce the observed distribution of $M_{\text{Z}}$ and $\dot{M}_{\text{Z}}$ for known polluted white dwarfs \citep{Wyatt2014,Turner2019}. These models have a distribution of parent body masses, parametrized by a power law slope and maximum mass. Two regimes are identified, either where a single accretion event dominates the photospheric abundances, or where multiple events contribute. The boundary between the two regimes depends, amongst other things, on the maximum planetesimal mass, found in the study to be $m_{\text{max}}=3\times10^{24}$\,g. For 2216$-$657, where a single event is implicated, a lower limit can be placed on the maximum planetesimal mass by extrapolating back to find $M_{\text{Z}}=10^{23}$\,g, which is consistent with the model. Identifying more stars in the decreasing phase may allow the model parameters to be better constrained.

\section{Summary}
\label{sectionSummary}

The study of atmospheric pollution at white dwarfs is a powerful and unique method for examining the bulk composition of exoplanetary objects. Deep, high-resolution spectroscopy gives access to multiple metal species, including the dominant rock-forming elements. This study presents observations of nine stars, and an analysis based on abundance ratios and comparison to solar system objects. The numerous sources of uncertainty are discussed, and results from the literature are used to show that observational and modelling uncertainties are sufficient to affect the outcome of a test for consistency with solar system objects. The results of such tests should therefore be treated as indicative rather than definitive.

Across the sample, the accreted material appears to be volatile-poor and rocky, as seen in the majority of systems studied to date. The stars display a range of compositions, some of which suggest an origin in a differentiated body, including one consistent with accretion of a rocky planetesimal with a small Fe--Ni core. Unusual ratios are observed: relative to bulk Earth, the material at two stars appears strongly enhanced in Na, while at another Ca is heavily depleted, and one system shows an enhancement in Mg attributed to differential sinking of elements since cessation of accretion around 5\,Myr ago.

The over-abundance of Na is investigated. Contamination by interstellar lines is unlikely, and the few data available do not implicate intrinsic variation in stellar chemistry as the cause. The slow-sinking Na may be left over from past pollution events, whose signatures have otherwise been obscured by ongoing accretion. Alternatively, as Na is relatively volatile, it may trace the local conditions (and therefore the location) in which the parent body of the accreted material formed. Abundances of other moderately volatile elements (e.g.~P and S), which are difficult to obtain with ground-based spectroscopy, will be required to test this hypothesis.

This study demonstrates the value of determining abundances for multiple elements. Looking to the future, high quality optical spectra that reveal the major rock-forming elements, ideally complemented by ultraviolet spectra to provide access to the volatiles (e.g. C, N, O), will be important. These data will enable detailed comparison with solar system objects, and confident identification of systems in the decreasing phase. As the number of stars surveyed increases, the uncertainties of individual objects become less important: population statistics can be compared with simulation outcomes, allowing issues such as the wide range of Na abundances to be addressed.

\section*{Acknowledgements}

The authors are grateful to the anonymous referee for a thorough review that led to significant improvements, and to Larry Nittler who kindly made available his database of meteorite measurements. This work is based on observations collected at the European Southern Observatory under ESO programme 087.D-0858. Some of the data presented herein were obtained at the W.~M.~Keck Observatory from telescope time allocated to NASA. This research is based on observations made with the NASA/ESA \textit{Hubble Space Telescope}, which is operated by the Association of Universities for Research in Astronomy, Inc., under NASA contract NAS~5-26555. These observations are associated with \textit{HST} program~14597. AS acknowledges support from a Science and Technology Facilities Council (STFC) studentship. JF acknowledges support from STFC grant ST/R000476/1. MH acknowledges funding from the European Union's Horizon~2020 research and innovation programme no.~677706 (WD3D). SGP acknowledges the support of an STFC Ernest Rutherford Fellowship. PWC and SR were supported by a NASA Keck PI Data Award, administered by the NASA Exoplanet Science Institute. BTG acknowledges support from STFC (ST/P000495/1).

This work has made use of: the SIMBAD database, operated at CDS, Strasbourg, France; the VALD database, operated at Uppsala University, the Institute of Astronomy RAS in Moscow, and the University of Vienna; data from the European Space Agency (ESA) mission
{\it Gaia} (\url{https://www.cosmos.esa.int/gaia}), processed by the {\it Gaia}
Data Processing and Analysis Consortium (DPAC,
\url{https://www.cosmos.esa.int/web/gaia/dpac/consortium}); the Pan-STARRS1 Surveys; and SkyMapper.




\bibliographystyle{mnras}
\bibliography{X-shooter_p87} 




\appendix

\section{\texorpdfstring{The $\upchi^{2}$ analysis and its interpretation}{The chi-squared analysis and its interpretation}}
\label{appendixChiSquared}

The widespread use of $\upchi^{2}$ in astronomy has caused some concern \citep{Andrae2010}, as it is only valid if the uncertainties are Gaussian. This is clearly not the case if logarithmic abundances with large Gaussian uncertainties are converted to mass fractions. However, a transform of the data can improve compliance with this requirement.

\subsection{The logit transform}
\label{subsectionLogit}
Proportional data are often encountered in the biological sciences, and methods for their analysis have received commensurate scrutiny, yielding a recommendation that the logit function (Eqn.~\ref{equationLogit}) be applied before analysis \citep{Warton2011}. 

The benefit of this transform is now demonstrated with simulated data. Abundances of the material polluting two hypothetical white dwarfs are assumed to be determined by an observer, subject only to Gaussian measurement uncertainties of 0.2~dex. In both cases, the material is a mix of only three minerals: a feldspar CaAl\textsubscript{2}Si\textsubscript{2}O\textsubscript{2}, olivine (FeMg)\textsubscript{2}SiO\textsubscript{8}, and an Fe--Ni alloy in the same ratio as the core of the Earth. The proportions of these minerals vary between the stars. One star is polluted with material having a feldspar\,:\,olivine\,:\,Fe--Ni ratio of 100\,:\,1\,:\,$x$, while the ratio at the other star is 1\,:\,100\,:\,$x$. The proportion of Fe--Ni alloy is varied from $x=0.1$ to $x=10\,000$ across a grid of simulated observations, identically for each star, effectively providing a constant background against which the feldspar and olivine vary. Ideally, the outcome of a consistency test should not be strongly influenced by the Fe--Ni concentration.

The $\upchi^{2}$~test is then used to determine whether the measurements are consistent with the material at each star having the same composition. The test is repeated under four separate conditions: using the computed mass fractions either (a)~directly, or (b)~logit-transformed, and assuming that either (i)~only Ca, Mg, and Fe have been detected, or (ii)~all seven polluting elements are detected. The test results are normalised to the 95~per~cent confidence level for the appropriate number of degrees of freedom (either~two or~six here; Eqn.~\ref{equationChiSquared}), so that a value of 1.0 or below in any given experiment indicates the two sets of measurements are consistent.

\begin{figure}
 \includegraphics[width=\columnwidth]{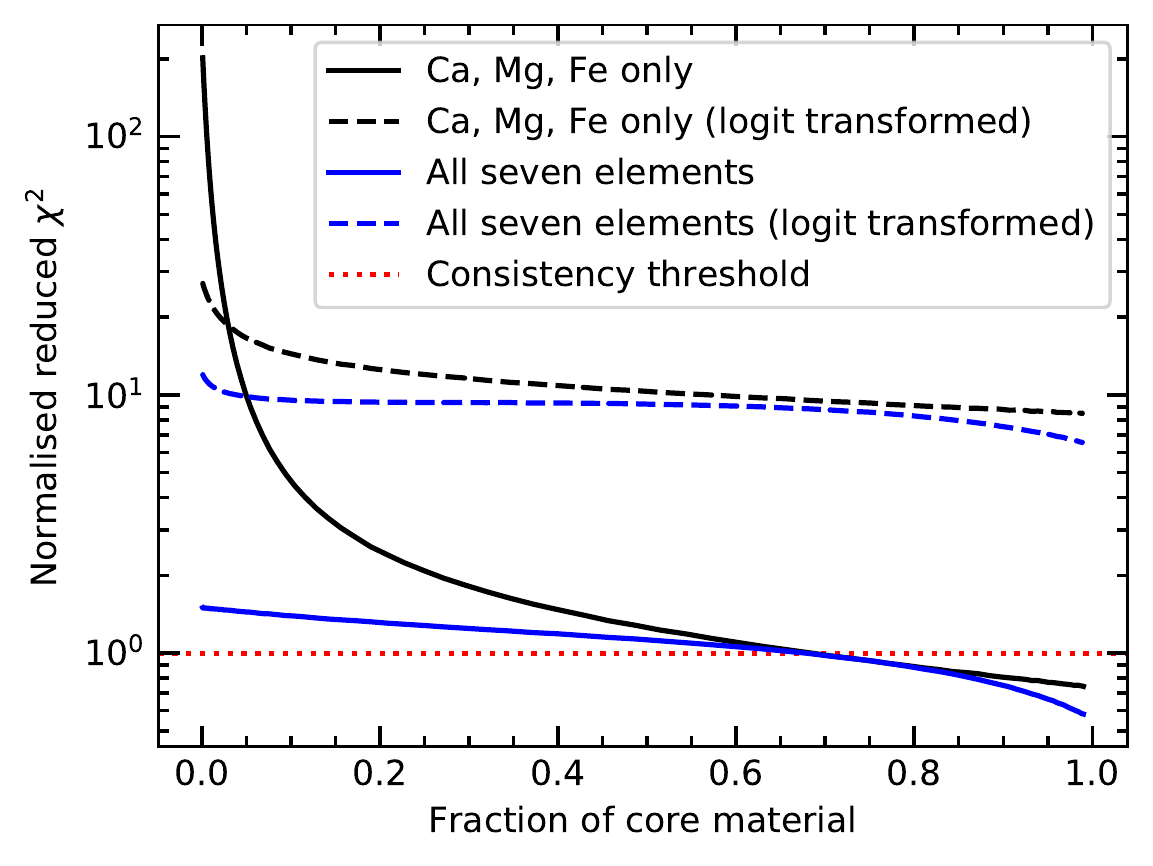}
 \caption{Results of the simulated experiment described in the text, showing that the logit transform improves the $\upchi^{2}$~consistency test.}
 \label{figureChiSquaredLogitDemo}
\end{figure}

The outcomes of this simulated analysis are shown in Fig.~\ref{figureChiSquaredLogitDemo}. The normalised~$\upchi^{2}$ is plotted for each experiment, with the dotted red line showing the consistency threshold. Despite the ratio between feldspar and olivine  differing by two orders of magnitude in the simulated materials, the standard $\upchi^{2}$~test fails to recognise that difference once Fe--Ni\,:\,$\text{feldspar}+\text{olivine}>2.2$ in each case. However, applying the logit transform reduces the sensitivity of the test to the absolute abundance of any given element, so that the test result is less affected by the level of the Fe--Ni background. Thus, it becomes more robust against false positives, i.e. materials are less likely to be identified as consistent with each other when there are significant differences between them.

\subsection{Application to real data}
\label{subsectionLogitApplication}
If the data presented in this paper are analysed without the logit transform, the results become more difficult to interpret. Most of the stars show consistency deeper into the decreasing phase, and with more comparison objects. This is especially true for 2105$-820$, where the polluting material becomes consistent with many irons and achondrites, well beyond the post-accretion times that are disfavoured due to upper limits, and even where the mass of accreted material would have exceeded that of the entire mixing layer. Note that abundance upper limits are \textit{not} used in the $\upchi^{2}$ analysis, so it is encouraging that the logit transform improves compliance with these limits.

A similar narrowing of the consistency results is seen using an example from the literature. Abundances published for the H-dominated white dwarf G29-38 are re-analysed using the methods presented here, with and without the logit transform \citep{Xu2014}. Ten elements are used, following the original study by including S and Ni at their upper limits. Accretion is assumed to be in a steady state, as sinking times are weeks to months. Bulk Earth and meteorite groups of the same types represented in the original study are found to be consistent with the accreted material. However, consistency is maintained well into the decreasing phase in some cases, beyond the limit imposed by the size of the mixing layer. Removing S and Ni from the analysis brings even more meteorite groups into consistency. The original study made comparisons against individual meteorite measurements that do not have abundance uncertainties, and most representatives of each meteorite class were found to be inconsistent. The analysis here groups meteorites of each class together and includes the spread of the distribution in the $\upchi^{2}$~analysis, so more classes are found to be consistent. However, when the logit transform is applied, the set of consistent groups narrows to just five, including those identified in the original study. Also, consistency no longer extends so far into the decreasing phase. The analysis is more robust as it considers meteorite groups and the spreads in their abundances, rather than relying on individual objects that cross the consistency threshold while others in their class do not. Given the uncertainties involved, seeking to identify best-fit blends of meteorites may be overfitting the data.

\subsection{Limitations on interpretation}
\label{subsectionErrors}
A strict interpretation of the $\upchi^{2}$ results requires well-quantified uncertainties, which are not straightforward to achieve for white dwarf abundances. There are many potential sources of error in the observations and analyses. For example, there is a linear relationship between the error in the oscillator strength for a line and the error in an abundance determined from that line. A recent study on the quality of publicly available atomic line data found transitions where literature values of $\log{(gf)}$ disagreed by up to 2~dex \citep{Laverick2018}. While such a large error could be easily identified and excluded, smaller discrepancies may remain. Thus, where multiple lines for a given species are present in a spectrum, the best fit to each may suggest different abundances. It must be hoped that such errors will cancel to first order, but there can be no guarantees. Likewise, the comparison samples also have their own uncertainties. Meteoritic studies work with materials that may have been processed during their journey through the atmosphere, have potentially been contaminated and weathered before recovery, and whose parent body is difficult to identify with confidence. Even the composition of bulk Earth is not yet settled: the abundance of C, for example, is constrained only within a factor of two \citep{Wang2018}.

\begin{figure}
 \includegraphics[width=\columnwidth]{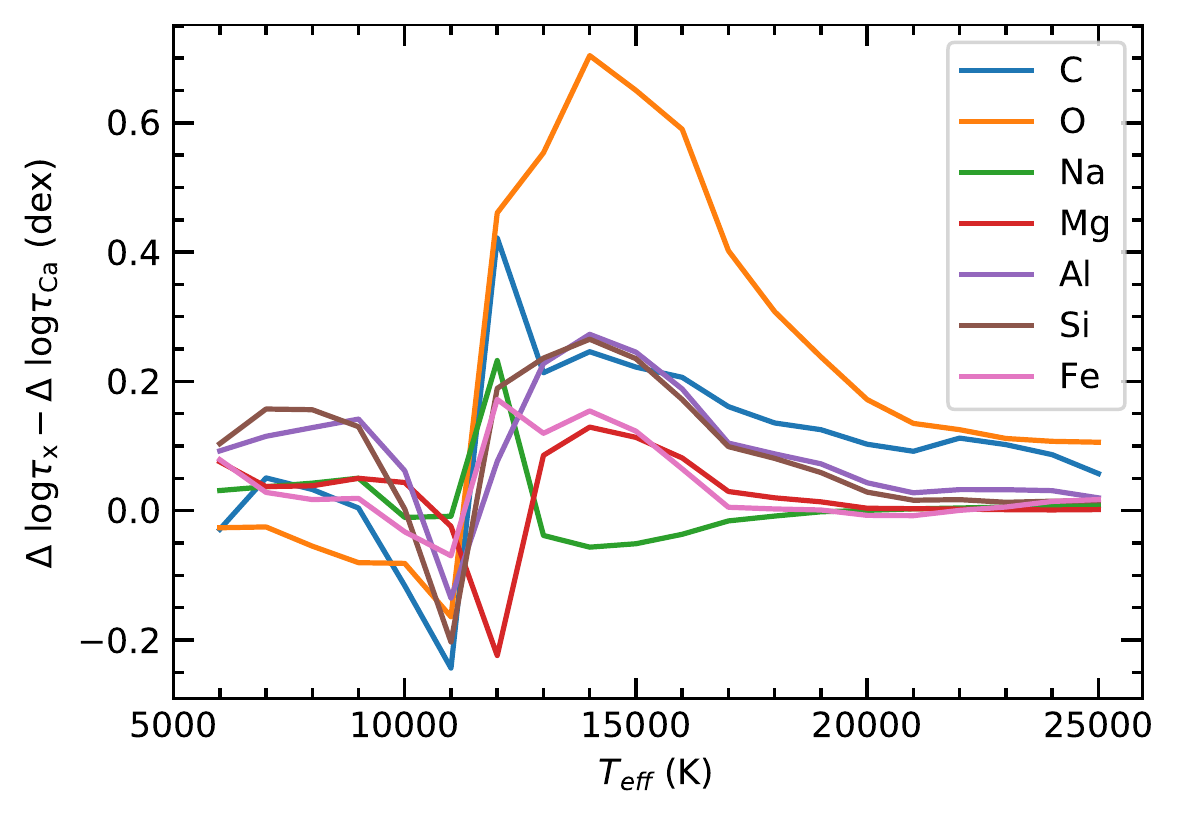}
 \caption{Discrepancy in sinking time-scales between Montreal and Koester models for H-dominated stars at $\log{g}=8.0$, normalised to Ca.}
 \label{figureSinkingTimes}
\end{figure}

Abundance determination is model-dependent, but the systematic errors can be investigated because multiple white dwarf atmosphere models have been developed by independent groups. For example, the Montreal group have calculated and made public\footnote{\href{http://dev.montrealwhitedwarfdatabase.org/evolution.html}{dev.\hspace{0pt}montreal\hspace{0pt}white\hspace{0pt}dwarf\hspace{0pt}database.\hspace{0pt}org/\hspace{0pt}evolution.html}} sinking times \citep{Fontaine2015,Dufour2017MWDD}. There is some disagreement with the model used here in the depth of the mixing layer, as well as the magnitude of sinking times, by factors of up to 85 for H-dominated stars and six for He-dominated stars. The spread of the disagreement among elements at a given point in $T_{\text{eff}}$--$\log{g}$ space is much smaller, typically only a factor of 1.2, as illustrated in Fig.~\ref{figureSinkingTimes}. The exception is O, where the difference can reach up to six times that of other elements in H-dominated stars around 14\,000~K. Further comparisons between models can be found in the literature \citep{Bauer2019, Cunningham2019}.

To demonstrate the impact of the less easily quantified uncertainties, a comparison is made between two published reports on the same stars \citep{Koester2011,Hollands2017abundances}. In the earlier study, cool white dwarfs were identified and analysed using SDSS spectra. The later study repeated the exercise on a larger scale, with updated SDSS data, new observations of some of the stars, and improved modelling of some spectral lines. Measurements of photospheric Ca, Mg, and Fe are available for 25~stars common to both studies.

\begin{figure}
 \includegraphics[width=\columnwidth]{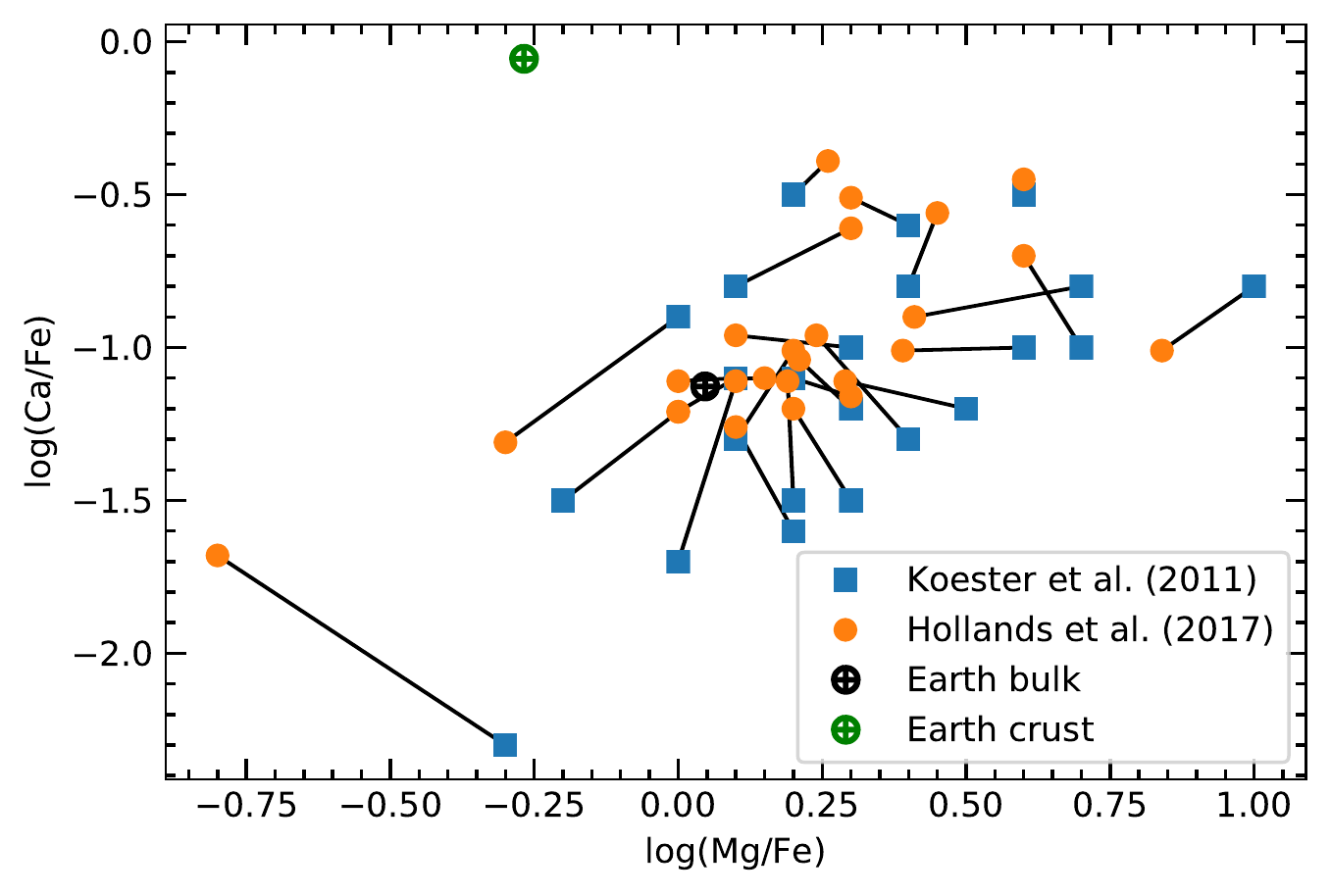}
 \caption{Abundance ratios of stars common to two similar studies, illustrating the scale of measurement and systematic errors. Earth abundances are shown for comparison.}
 \label{figureKoesterVsHollands}
\end{figure}

Photospheric number abundances from each study are plotted in Fig.~\ref{figureKoesterVsHollands}. The differences in $\ratio{Ca}{Fe}$ ratios are never as large as the difference between bulk Earth and the crust, but the same is not true of the differences in $\ratio{Mg}{Fe}$ ratios. However, the mean differences in ratios across the sample are small: $\langle\upDelta\log{(\ratio{Mg}{Fe})}\rangle=0.07$, and $\langle\upDelta\log{(\ratio{Ca}{Fe})}\rangle=0.14$ . The differences in $\log{(\ratio{Ca}{Fe})}$ show a correlation with $T_{\text{eff}}$ (correlation coefficient~$\uprho=0.54$) but the differences in $\log{(\ratio{Mg}{Fe})}$ do not ($\uprho=0.11$). The stars in these studies all have sinking times longer than $10^{5}$\,yr, so variation in photospheric abundances between observations is not expected. The differences are therefore likely attributable to systematic errors, arising in observations, data reduction, and modelling. New data increased the wavelength coverage, and therefore the number of lines available for fitting, and the model had evolved during the time between the studies.

Applying the $\upchi^{2}$~analysis to both sets of data leads to different conclusions about the consistency of the material at each star with the comparison data. Of the 25~stars, only one remains consistent with the same set of comparisons across both studies. In most cases, shifts into or out of consistency occur for several comparisons, while for four stars more than half the comparisons cross the consistency threshold. The results of the $\upchi^{2}$ consistency test should therefore be treated as indicative, and not strictly interpreted. Likewise, it may be premature to take an apparent change in abundances at a star as evidence of variation in the composition of accreted material, if those abundances were determined in heterogeneous analyses.


\bsp	
\label{lastpage}
\end{document}